\documentclass[aps,eqsecnum,superscriptaddress,preprintnumbers]{revtex4}

\newcommand{\beq}{\begin{equation}}
\newcommand{\eeq}{\end{equation}}
\newcommand{\beqa}{\begin{eqnarray}}
\newcommand{\eeqa}{\end{eqnarray}}
\newcommand{\lslash}[1]{#1\llap/}

\newcommand{\limQzero}{\vec Q\rightarrow 0}

\begin{document}

\preprint{hep-ph/0208146}
\title{
Field theory of the photon self-energy in a medium with a 
magnetic field and the Faraday effect
}
\author{Juan Carlos D'Olivo}
\affiliation{Instituto de Ciencias Nucleares\\
Universidad Nacional Aut\'onoma de M\'exico\\
Circuito Exterior, C.U., A. Postal 70-543\\
04510 Mexico DF, Mexico}

\author{Jos\'e F. Nieves}
\affiliation{Laboratory of Theoretical Physics, 
Department of Physics, P.O. Box 23343\\
University of Puerto Rico, 
R\'{\i}o Piedras, Puerto Rico 00931-3343}

\author{Sarira Sahu}
\affiliation{Instituto de Ciencias Nucleares\\
Universidad Nacional Aut\'onoma de M\'exico\\
Circuito Exterior, C.U., A. Postal 70-543\\
04510 Mexico DF, Mexico}

\date{August 2002}

\begin{abstract}

A convenient and general decomposition of the photon self-energy
in a magnetized, but otherwise isotropic, medium is given in terms
of the minimal set of tensors consistent with the transversality
condition. As we show, the self-energy in such a medium is
completely parametrized in terms of nine independent form factors,
and they reduce to three in the long wavelength limit. We consider
in detail an electron gas with a background magnetic field, and
using finite temperature field theory methods, we obtain the
one-loop formulas for the form factors, which are exact to all
orders in the magnetic field. Explicit results are derived for a
variety of physical conditions. In the appropriate limits, we recover
the well-known semi-classical results for the photon dispersion
relations and the Faraday effect. In more general cases, where the
semi-classical treatment or the linear approximation (weak field
limit) are not applicable, our formulas provide a consistent and
systematic way for computing the self-energy form factors and,
from them, the photon dispersion relations.

\end{abstract}

\maketitle
%
% sec 1
%
\section{Introduction}

The propagation of photons in a material medium in the presence of
a magnetic field gives rise to many observable effects. For
example, in astrophysical contexts, the fact that the photons with
different polarizations have different dispersion relations
(birefringence), leads to the Faraday rotation effects that have
been observed for various astrophysical
objects\cite{faradayeffects1,faradayeffects2,faradayeffects3,
faradayeffects4,faradayeffects5,faradayeffects6}.
Recently\cite{melnikovetal}, the observation of  birefringence
effects have also been reported in an experiment that studied the
millisecond pulsations of solar radio emission. In a different
context, in another recent laboratory
experiment\cite{verbiestetal}, an effect that mimics the Faraday
rotation has been found. In this case, using two input beams (a
control and a probe) it was shown that a Faraday-like effect is
induced, for both circular and linear polarizations of the control
beam. From a theoretical point of view, the common theme in all
these situations is the propagation of a photon in matter in the
presence of an external field, being a constant magnetic field in
one case, or a plane wave electromagnetic field in the other.

The subject of the propagation of photons in magnetized plasmas
is of course a well studied one, and it is amply covered in standard
astrophysics and plasma physics textbooks\cite{ishimaru,llphyskin}.
The traditional approach to this subject has been
based on the equations of classical kinetic theory and
related semiclassical methods. On the other hand, it is
commonly accepted that the field-theoretical
methods are generally useful in this kind of problem,
since they are applicable to a wider range of physical situations
for which the semiclassical methods breakdown. Such situations include,
for example, matter under extreme conditions, such as highly
relativistic and/or highly degenerate.

For the weak field limit a field-theoretical treatment has been
given in Ref.\ \cite{paletal}. The assumption there is that the
photon self-energy can be expanded as a power series in the
external magnetic field and that the linear terms are the relevant
one. There exist physical situations that lie outside of the scope
of such approximation and require a more general treatment. For
example, the arguments for the existence of super critical fields
($\ge 4.4\times 10^{13}$ G) in nature have grown stronger recently
with the observations of few Soft Gamma Rays Repeaters (SGR) and
Anomalous X-Rays Pulsars (AXP), which are very likely
magnetars\cite{magnetar1,magnetar2}, that is, isolated neutron
stars with surface magnetic fields of order $10^{14}-10^{15}$ G. A
model for extragalactic gamma rays burst in terms of merger of
massive binary stars suggests also that magnetic fields up to the
order of $10^{17}$ G\cite{narayan} may exist. In the context of
the Early Universe, very large magnetic fields ($10^{23}$ G) may
be generated\cite{vachaspati} during the electroweak phase
transition due to gradients in the Higgs field. Thus, there are
environments of interest - that involve, in addition to matter in
extreme relativistic and/or degenerate conditions, strong magnetic
fields - for which neither the semiclassical methods nor the weak
field approach are directly applicable.

Our purpose here is to have another look at this subject, keeping
the above points in mind. To this end, we consider the photon
propagation in an electron gas in the presence of a constant
external magnetic field. We derive the general decomposition of
the photon self-energy in such a medium, in terms of the minimal
set of tensors consistent with isotropy and the transversality
condition. We show that the self-energy is completely parametrized
in terms of nine independent form factors, and that in the long
wavelength limit only three are independent, with the rest being
expressed in terms of them. Using the (real-time) finite
temperature field theory method, we obtain the one-loop
expressions for the form factors, expressed as integrals over the
distribution functions of the particles in the background. Those
formulas are valid to all orders in the magnetic field and are
explicitly evaluated for a variety of conditions of physical
interest. As an application we determine the photon dispersion
relations and calculate the Faraday rotation 
for plane polarized light, for various cases.  
They reproduce the well-known semi-classical results
when the appropriate limits are taken, but remain valid for
more general situations.  

In Section\ \ref{sec:kinematics} we give the general
decomposition of the photon self-energy in terms of the
nine independent form factors. We also collect
there several kinematical relations that are useful
in later stages of the calculations. The one-loop formulas
for the photon polarization tensor are derived in Section\ \ref{sec:oneloop}.
These are used in Section\ \ref{sec:disprel} to discuss the
dispersion relations, focusing on the long wavelength limit
as a special case. The explicit formulas for the independent
form factors are given in Section\ \ref{sec:explicitformulas}
for different possible conditions of the background electron gas,
and for various regimes of interest including the \emph{low frequency regime}
and the \emph{weak-field (linear) limit}.
Our conclusions are summarized in Section\ \ref{sec:conclusions}, while
in Appendix\ \ref{app:propagator} we summarize the conventions
that we use regarding the Schwinger formula for the electron
propagator in an external magnetic field, and
Appendix\ \ref{app:J0} contains the derivation of an integral
formula used in the calculation. Some of the details of the
derivation of the low-frequency formulas, and the weak-field
formulas, for the self-energy form factors are shown
in appendices \ref{app:lowfrequencyregime} and \ref{app:weakfieldlimit},
respectively.
%
% sec2
%
\section{Kinematics}
\label{sec:kinematics}

\subsection{General decomposition of the polarization tensor}
In the absence of the magnetic field,
the photon self-energy $\pi_{\mu\nu}$ depends in general
on the photon momentum $q^\mu$, and on the velocity four-vector of the
medium $u^\mu$. In the frame of reference in which the medium is at rest,
$u^\mu$ has components given by
\beq
u^\mu = (1,\vec 0) \,,
\eeq
and in that frame we write
\beq
q^\mu = (\omega,\vec Q) \,.
\eeq
In the presence of a magnetic field, but otherwise an isotropic medium,
$\pi_{\mu\nu}$ depends in addition on the vector $b^\mu$ that is determined
by the magnetic field. The vector $b^\mu$ is defined such that,
in the frame in which the medium is at rest,
\beq
b^\mu = (0,\hat b) \,,
\eeq
where we denote the magnetic field vector by
\beq
\vec B = B\hat b \,.
\eeq

For a given photon momentum vector
we define the unit vectors $\hat e_i$ ($i = 1,2,3$) by writing
\beq
\vec Q \equiv Q \hat e_3 \,,
\eeq
with $\hat e_{1,2}$ chosen such that
\beqa
\hat e_1\cdot \hat e_3 & = & e_1\cdot \hat e_3 = 0 \,,\nonumber\\
\hat e_2 & = & \hat e_3 \times \hat e_1 \,.
\eeqa
In addition,
for the problem that we are considering in the present work,
without loss of generality, we can choose the vectors
$\hat e_{1,2}$ such that $\hat b$ lies in
the $1,3$ plane. Thus we can write,
\beq
\hat b = \cos\theta\hat e_3 + \sin\theta\hat e_1 \,,
\eeq
where
\beq
\cos\theta = \hat Q\cdot\hat b \,.
\eeq

We now introduce the vectors
\beqa
X_3^\mu & = & u^\mu - \frac{(u\cdot q) q^\mu}{q^2} \,, \nonumber\\
X_2^\mu & = &
\epsilon^{\mu\alpha\beta\gamma}q_\alpha b_\beta u_\gamma \,, \nonumber\\
X_1^\mu & = & \epsilon^{\mu\alpha\beta\gamma}X_{2\alpha} q_\beta u_\gamma \,,
\eeqa
which satisfy
\beqa
q\cdot X_i & = & 0 \,, \nonumber\\
X_i\cdot X_j & = & 0 \qquad (i \not = j) \,.
\eeqa
Therefore, the $X_i^\mu$ form a basis of vectors orthogonal to
$q^\mu$, and the nine bilinear combinations
$X_i^\mu X_j^\nu$
form a complete set of tensors in terms of which the photon
self-energy can be decomposed. To exploit this more fully, it is
useful to define the (normalized) vectors
\beq
\epsilon^\mu_i(\vec Q) = \frac{X_i^\mu}{\sqrt{-X\cdot X}} \,,
\eeq
which satisfy
\beqa
\epsilon_i(\vec Q)\cdot q & = & 0 \,,\nonumber\\
\epsilon_i(\vec Q)\cdot\epsilon_j(\vec Q) & = & -\delta_{ij} \,,\nonumber\\
g_{\mu\nu} - \frac{q^\mu q^\nu}{q^2} & = &
-\sum_{i = 1,3}\epsilon_{\mu i}(\vec Q) \epsilon_{\nu j}(\vec Q)\,,
\eeqa
and which take the form
\beqa
\label{defepsilons}
\epsilon^\mu_1(\vec Q) & = & (0,\hat e_1) \,,\nonumber\\
\epsilon^\mu_2(\vec Q) & = & (0,\hat e_2) \,,\nonumber\\
\epsilon^\mu_3(\vec Q) & = & \frac{1}{\sqrt{q^2}}(Q,\omega\hat e_3)
\eeqa
in the rest frame of the medium. 

Whence, in the most general case,
the photon self-energy can be expressed as a linear combination of
the bilinears $\epsilon^\mu_i\epsilon^\nu_j$ in the form
\beq 
\label{gendecomp}
\pi^{(eff)}_{\mu\nu}(\omega,\vec{Q}) =
-\sum_{i,j}\pi^{(ij)}(\omega,\vec{Q}) \epsilon_{\mu i}(\vec Q)
\epsilon_{\nu j}(\vec Q) \,,
\eeq
involving the nine independent coefficient functions
$\pi^{(ij)}(\omega,\vec{Q})$. Furthermore, as will be seen in 
Section\ \ref{subsec:lwlimit}, in the long wavelength limit
all the $\pi^{(ij)}(\omega,\vec{Q})$ can be determined in terms
of just three independent functions [i.e., Eq.\ (\ref{pitildeQ0})].
This fact will be the
starting point in Section\ \ref{sec:disprel} to obtain the
polarization vectors and dispersion relations of the normal modes.

\subsection{Additional kinematic relations}
It will prove to be useful to introduce the following tensors that are
transverse to $u^\mu$,
\beqa 
\label{deftensors} 
Q_{\mu\nu} & = & -b_\mu b_\nu \,, \nonumber\\
R_{\mu\nu} & = & g_{\mu\nu} - u_\mu u_\nu - Q_{\mu\nu} \,, \nonumber\\
P_{\mu\nu} & = & i\epsilon_{\mu\nu\alpha\beta}b^\alpha u^\beta \,.
\eeqa
They have the orthogonality properties
\beq
QR = QP = 0 \,,
\eeq
and they satisfy the multiplication rules
\beqa
Q^2 & = & Q \,, \nonumber\\
R^2 & = & R \,, \nonumber\\
P^2 & = & R \,, \nonumber\\
RP & = & P \,,
\eeqa
as well as the normalization conditions
\beqa
Q^\mu{}_\mu & = & 1 \,, \nonumber\\
R^\mu{}_\mu & = & 2 \,.
\eeqa
In addition, we define
\beq
U_{\mu\nu} = u_\mu b_\nu - b_\mu u_\nu \,,
\eeq
which satisfies
\beqa
UR & = & RU \,, \nonumber\\
U^{\mu\lambda} U_{\lambda\nu} & = & u^\mu u_\nu - b^\mu b_\nu \,, \nonumber \\
Q^{\mu\nu}U_{\mu\alpha}U_{\nu\beta} & = & -u_\alpha u_\beta \nonumber\\
& = & g^\mu{}_{\nu} - R^\nu{}_{\mu} \,, \nonumber\\
\frac{1}{2}P^{\mu\nu}\epsilon_{\mu\nu\alpha\beta} & = & iU_{\alpha\beta}\,.
\eeqa
In any case, any four-vector $v^\mu$ can be decomposed according to
\beq
v^\mu = v^\mu_\parallel + v^\mu_\perp \,,
\eeq
where
\beqa
\label{vperppar}
v^\mu_\perp & = & R^\mu{}_\nu v^\nu  = (0,\vec v_\perp) \,,\nonumber\\[12pt]
v^\mu_\parallel & = & v^\mu - v^\mu_\perp  = (v^0,\vec v_\parallel)\,,
\eeqa
with $\vec v_\perp$ and $\vec v_\parallel$ being the three-dimensional
components of $\vec v$ that are perpendicular and parallel to $\vec b$,
respectively.
In particular, for any two such vectors,
\beqa
v_\perp\cdot v^\prime_\perp & = &
v\cdot v^\prime - (v\cdot u)(v^\prime\cdot u) + (v\cdot b)(v^\prime\cdot b)
\,, \nonumber\\
v_\parallel\cdot v^\prime_\parallel & = &
(v\cdot u)(v^\prime\cdot u) - (v\cdot b)(v^\prime\cdot b) \,.
\eeqa
%
%
% sec 3
%
\section{Photon self-energy}
\label{sec:oneloop}

In one-loop, the 11 element of the photon thermal self-energy is
given by
\beq \label{selfenergy} i\pi_{11\mu\nu}(\omega,\vec Q) =
-(-ie)^2\int \frac{d^4p}{(2\pi)^4} \mbox{Tr}\,\left[\gamma_\mu
iS_e(p+q)\gamma_\nu iS_e(p)\right], 
\eeq
where $S_e$ stands for the 11 component of the electron thermal
propagator. It can be decomposed in the form
\beq 
\label{Se} 
S_e(p) = S_{Fe}(p) + S_{Te}(p) \,, 
\eeq
where $S_{Fe}$ is the propagator in vacuum in the presence of the
magnetic field while $S_{Te}$ incorporates the effects of the
thermal background. As shown in Appendix\ \ref{app:propagator},
$S_F(p)$ can be written in the form
\beq 
\label{SFe} 
iS_F(p) = \int_0^{\infty}d\tau G(p,s)e^{i\tau\Phi(p,s) - \epsilon\tau} \,,
\eeq
where
\beqa 
\label{GPhi} 
\Phi(p,s) & = & p^2 - m^2_e +
\left(s^{-1}\tan(s) - 1\right)R_{\mu\nu} p^\mu p^\nu\,,\nonumber\\
G(p,s) & = & \lslash{V}_1(p,s) + i\gamma_5\lslash{V}_2(p,s) + m_e
+ m_e\tan(s)i\gamma_5\lslash{u}\lslash{b} \,, 
\eeqa
with
\beqa 
\label{V12} 
V_{1\mu}(p,s) & = & \sec^2(s)p_\mu +
\tan^2(s)\left[(p\cdot b)b_\mu - (p\cdot u)u_\mu\right] \,,\nonumber\\ 
V_{2\mu}(p,s) & = & \tan(s)\left[(p\cdot b)u_\mu - (p\cdot u)b_\mu\right]
\eeqa
and
\beq 
\label{vars} s = |e|B\tau \,. 
\eeq
Using the relation $\sec^2(s) = 1 + \tan^2(s)$, Eq.\ (\ref{V12})
can be rewritten as
\beqa 
\label{V12tensors} 
V_{1\mu}(p,s) & = & \left[g_{\mu\nu} +
\tan^2(s)R_{\mu\nu}\right] p^\nu \,,\nonumber\\ 
V_{2\mu}(p,s) & = &\tan(s)U_{\mu\nu}p^\nu \,. 
\eeqa
On the other hand, the thermal part is given by
\beq 
\label{STe} 
iS_{Te}(p) = -\eta_e(p)\int^{\infty}_{-\infty} d\tau G(p,s) 
e^{i\tau\Phi(p,s) - \epsilon|\tau|} \,,
\eeq
where
\beq 
\label{etae} 
\eta_e(p) = \theta(p\cdot u)f_e(p\cdot u) +
\theta(-p\cdot u)f_{\bar e}(-p\cdot u)\,, 
\eeq
with
\beqa 
f_e(x) & = & \frac{1}{e^{\beta(x - \mu_e)} + 1} \nonumber\\
f_{\bar e}(x) & = & \frac{1}{e^{\beta(x + \mu_e)} + 1} \,. 
\eeqa
Here $\beta$ stands for the inverse  temperature and $\mu_e$ the
electron chemical potential.

When Eq.\ (\ref{Se}) is substituted into Eq.\ (\ref{selfenergy}),
we obtain four different terms. Only those two that contain
one factor of $S_{Fe}$ and $S_{Te}$ each, contribute to the real
part of the self-energy, which we denote by
$\pi^{(eff)}_{\mu\nu}$, and is the quantity in which we are
interested. Thus
\beqa 
\pi^{(eff)}_{\mu\nu}(\omega,\vec Q) & = &
ie^2\int\frac{d^4p}{(2\pi)^4}\left\{
\eta_e(p)\int_{-\infty}^{\infty} d\tau e^{\lambda(p,\tau)}
\int_{0}^{\infty}  d\tau^\prime e^{\lambda(p + q,\tau^\prime)}
\mbox{Tr}\left[\gamma_\mu G(p+ q,\tau^\prime)\gamma_\nu
G(p,\tau)\right] \right. \nonumber\\ && \mbox{} + \left. \eta_e(p
+ q)\int_{0}^{\infty}d\tau e^{\lambda(p,\tau)}
\int_{-\infty}^{\infty}d\tau^\prime e^{\lambda(p + q,\tau^\prime)}
\mbox{Tr}\left[\gamma_\mu G(p + q,\tau^\prime)\gamma_\nu
G(p,\tau)\right] \right\}\,, 
\eeqa
where, to simplify the notation, we have defined
\beq 
\lambda(p,\tau) = i\tau\Phi(p,s) - \epsilon|\tau| \,. 
\eeq
In the term involving the factor $\eta_e(p + q)$, we make the
change of variables $p\rightarrow p - q$, and $\tau
\leftrightarrow \tau^\prime$. Then, using the cyclic property of
the trace, we obtain
\beq 
\label{pieff1} 
\pi^{(eff)}_{\mu\nu}(\omega,\vec Q) =
4ie^2\int\frac{d^4p}{(2\pi)^4}\eta_e(p) \int_{-\infty}^{\infty}
d\tau e^{\lambda(p,\tau)} \left\{ \int_{0}^{\infty}  d\tau^\prime
\left[ e^{\lambda(p + q,\tau^\prime)}L_{\mu\nu}(p,q) +
e^{\lambda(p - q,\tau^\prime)}L_{\nu\mu}(p,-q)\right]\right\} \,,
\eeq
where
\beq 
L_{\mu\nu}(p,q) = \frac{1}{4}\mbox{Tr}\left[\gamma_\mu
G(p+q,\tau^\prime) \gamma_\nu G(p,\tau)\right] \,. 
\eeq
Using the expression for $G$ given in Eq.\ (\ref{GPhi}), by
straightforward evaluation of the trace we obtain
\beq 
L_{\mu\nu} = L^{(s)}_{\mu\nu} + L^{(a)}_{\mu\nu}\,, 
\eeq
where
\beqa 
L^{(s)}_{\mu\nu}(p,q) & = & V_{1\mu}V^\prime_{1\nu} +
V^\prime_{1\mu}V_{1\nu} - V_1\cdot V^\prime_1 g_{\mu\nu} -
V_{2\mu}V^\prime_{2\nu} - V^\prime_{2\mu}V_{2\nu} + V_2\cdot
V^\prime_2 g_{\mu\nu} \,,\nonumber\\ 
&& \mbox{} + m^2_e g_{\mu\nu} +
m^2_e\tan(s)\tan(s^\prime)\left(g_{\mu\nu} - 2u_\mu u_\nu + 2
b_\mu b_\nu\right) \,,\nonumber\\
L^{(a)}_{\mu\nu}(p,q) & = &
\epsilon_{\mu\nu\alpha\beta}V^{\prime\alpha}_1 V^{\beta}_2 +
\epsilon_{\mu\nu\alpha\beta}V^{\prime\alpha}_2 V^{\beta}_1 +
m^2_e\left[\tan(s) -
\tan(s^\prime)\right]\epsilon_{\mu\nu\alpha\beta}u^\alpha b^\beta \,, 
\eeqa
with the convention that $\epsilon^{0123} = 1$, and using the
notation $s^\prime = |e|B\tau^\prime$, $V^\prime_{1\mu} =
V_{1\mu}(p + q,s^\prime)$, and similarly for $V^\prime_{2\mu}$.
Using Eq.\ (\ref{V12tensors}), $L^{(s,a)}_{\mu\nu}$ can be
expressed in the form
\beqa 
\label{Lsaexp} 
L^{(s)}_{\mu\nu}(p,q) & = & 2m^2_e
\tan(s)\tan(s^\prime) R_{\mu\nu} + m^2_e\left[1 -
\tan(s)\tan(s^\prime)\right] g_{\mu\nu} - g_{\mu\nu}\left[1 +
\tan(s)\tan(s^\prime)\right](p^2 + p\cdot q)\nonumber\\ & &
\mbox{} - g_{\mu\nu}\left[ \tan^2(s) + \tan^2(s^\prime) -
\tan(s)\tan(s^\prime) + \tan^2(s)\tan^2(s^\prime)
\right](p^2_\perp + p_\perp\cdot q_\perp) \nonumber\\ 
& & \mbox{}
+ \left\{\left[g_{\mu\alpha} + \tan^2(s) R_{\mu\alpha}\right]
\left[g_{\nu\beta} + \tan^2(s^\prime) R_{\nu\beta}\right] -
\tan(s)\tan(s^\prime) U_{\mu\alpha}U_{\nu\beta} +
(\mu\leftrightarrow\nu) \right\}p^\alpha(p + q)^\beta \,,\nonumber\\
L^{(a)}_{\mu\nu}(p,q) & = & \epsilon_{\mu\nu\alpha\beta}\left[
\tan(s)g^{\alpha\lambda}U^{\beta\rho} +
\tan(s^\prime)U^{\alpha\lambda}g^{\beta\rho} +
\tan(s)\tan^2(s^\prime)R^{\alpha\lambda}U^{\beta\rho} +
\tan(s^\prime)\tan^2(s)U^{\alpha\lambda}R^{\beta\rho}\right]
p_\lambda(p + q)_\rho \nonumber\\ &&\mbox{} + m^2_e(\tan(s) -
\tan(s^\prime)) \epsilon_{\mu\nu\alpha\beta}u^\alpha b^\beta \,,
\eeqa
where $p^\mu_\perp$ and $q^\mu_\perp$ are defined according to
Eq.\ (\ref{vperppar}). It is seen from Eq.\ (\ref{Lsaexp}) that
\beq 
L^{(s,a)}_{\mu\nu}(-p,q) = L^{(s,a)}_{\mu\nu}(p,-q) \,. 
\eeq
Using this and the fact that $\lambda(-p,s) = \lambda(p,s)$, 
Eq.\ (\ref{pieff1}) can be written in the form
\beqa 
\label{pimunuefffinal} 
\pi^{(eff)}_{\mu\nu}(\omega,\vec Q) &
= & 4ie^2\int_{p\cdot u \geq 0}\frac{d^4p}{(2\pi)^4}
\int_{-\infty}^{\infty} d\tau e^{\lambda(p,\tau)}
\int_{0}^{\infty}  d\tau^\prime \left\{\left[ e^{\lambda(p +
q,\tau^\prime)} L^{(s)}_{\mu\nu}(p,q)[f_e(p\cdot u) + f_{\bar
e}(p\cdot u)] + (q\rightarrow -q)\right]\right. \nonumber\\ &&
\mbox{} + \left.\left[e^{\lambda(p +
q,\tau^\prime)}L^{(a)}_{\mu\nu}(p,q) [f_e(p\cdot u) - f_{\bar
e}(p\cdot u)] - (q\rightarrow -q)\right]\right\} \,. 
\eeqa

The integrals over $\tau,\tau^\prime$ can be expressed in the form
\beqa 
\label{intLsaexp} 
\int_{-\infty}^{\infty} d\tau
e^{\lambda(p,\tau)} \int_{0}^{\infty}  d\tau^\prime e^{\lambda(p +
q,\tau^\prime)}L^{(s)}_{\mu\nu}(p,q) & = & -2m^2_e K_1(p)J_1(p +
q) R_{\mu\nu} + m^2_e g_{\mu\nu}\left[K_0(p)J_0(p + q) +
K_1(p)J_1(p + q)\right]\nonumber\\ &&\mbox{} -
g_{\mu\nu}\left[K_0(p)J_0(p + q) - K_1(p)J_1(p + q)\right](p^2 +
p\cdot q) \nonumber\\ 
& & \mbox{} - g_{\mu\nu}\left[ -K_2(p)J_0(p
+ q) - K_0(p)J_2(p + q)\right.\nonumber\\ &&\mbox{} \left. +
K_1(p)J_1(p + q) + K_2(p)J_2(p + q) \right](p^2_\perp +
p_\perp\cdot q_\perp) \nonumber\\ & & \mbox{} + \left[
g_{\mu\alpha}g_{\nu\beta}K_0(p)J_0(p + q) -
R_{\mu\alpha}g_{\nu\beta}K_2(p)J_0(p + q)\right.\nonumber\\
&&\mbox{} \left. - g_{\mu\alpha}R_{\nu\beta}K_0(p)J_2(p + q) +
R_{\mu\alpha}R_{\nu\beta}K_2(p)J_2(p + q)\right.\nonumber\\
&&\mbox{} + \left.U_{\mu\alpha}U_{\nu\beta}K_1(p)J_1(p + q)  +
(\mu\leftrightarrow\nu) \right]p^\alpha(p + q)^\beta \,,\nonumber\\
\int_{-\infty}^{\infty} d\tau e^{\lambda(p,\tau)}
\int_{0}^{\infty}  d\tau^\prime e^{\lambda(p +
q,\tau^\prime)}L^{(a)}_{\mu\nu}(p,q) & = &
i\epsilon_{\mu\nu\alpha\beta}\left[ K_1(p)J_0(p +
q)g^{\alpha\lambda}U^{\beta\rho} + K_0(p)J_1(p +
q)U^{\alpha\lambda}g^{\beta\rho}\right.\nonumber\\ 
&&\mbox{} -
\left.K_1(p)J_2(p + q)R^{\alpha\lambda}U^{\beta\rho} - 
K_2(p)J_1(p + q)U^{\alpha\lambda}R^{\beta\rho}\right] p_\lambda(p +
q)_\rho\nonumber\\ &&\mbox{} + m^2_e
i\epsilon_{\mu\nu\alpha\beta}u^\alpha b^\beta \left[K_1(p)J_0(p + q) 
- K_0(p)J_1(p + q)\right] \,, 
\eeqa
where
\beqa 
\label{Ji} 
J_n(p) & = & \frac{(-i)^n}{|e|B}\int_0^\infty ds
e^{is\psi_p}e^{-ia_p\tan(s)} \tan^n(s)\,,
%J_0(p) & = & \frac{1}{|e|B}\int_0^\infty ds e^{is\psi_p}e^{-ia_p\tan(s)}
%\nonumber\\
%J_1(p) & = & \frac{-i}{|e|B}\int_0^\infty ds e^{is\psi_p}e^{-ia_p\tan(s)}
%\tan(s)\nonumber\\
%J_2(p) & = & \frac{1}{|e|B}\int_0^\infty ds e^{is\psi_p}e^{-ia_p\tan(s)}
%\tan^2(s)\,,
\eeqa
with
\beqa 
a_p & \equiv & \frac{-p^2_\perp}{|e|B} = 
\frac{\vec p^{\;2}_\perp}{|e|B}\,,\nonumber\\ 
\psi_p & \equiv &
\frac{1}{|e|B}\left(p^2_{\parallel} - m^2_e + i\epsilon\right) =
\frac{1}{|e|B}\left(p_0^2 - \vec p^{\;2}_{\parallel} - 
m^2_e + i\epsilon\right)\,.
\eeqa
The $K_n$ are defined by the same integrals as in Eq.\ (\ref{Ji}), 
but with the limits
of integration being $-\infty < s < \infty$ instead. Therefore,
the following relations are easily obtained
\beq 
\label{Ki}
%K_n(p) = J_n(p) + J^\ast_n(p) \qquad (n = 1,2,3) \,.
K_n(p) = J_n(p) + J^\ast_n(p) \,. 
\eeq
Furthermore, from Eq.\ (\ref{Ji}) it is easily seen that
\beq 
\label{J2} 
J_{n + 1}(p) = \frac{\partial J_n(p)}{\partial a_p} \,, 
\eeq
and therefore, only $J_0$ needs to be evaluated explicitly. As
shown in Appendix\ \ref{app:J0}, it can be expressed in the form
\begin{equation}
\label{J1final} 
J_0(p) = \frac{i}{|e|B}\sum_{\ell = 0}^\infty
\frac{D_\ell(a_p)}{\psi_p - 2\ell} \,,
\end{equation}
where the $D_n(a_p)$ are given in terms of the Laguerre
polynomials by
%
%\begin{equation}
%\label{Dn}
%D_n(a) \equiv (-1)^n e^{-a}\times
%\left\{
%\begin{array}{ll}
%L_0(2a) & \mbox{for $n = 0$} \\
%L_n(2a) - L_{n - 1}(2a) & \mbox{for $n \ge 1$} \,.
%\end{array}
%\right.
%\end{equation}
%
\begin{equation}
\label{Dn} 
D_n(a) \equiv (-1)^n e^{-a} \left[L_n(2a) - L_{n -1}(2a)\right] \,,
\end{equation}
with the convention
\beq 
L_{-1} \equiv 0\,. 
\eeq
{}From the relations in Eq.\ (\ref{J2})
%and (\ref{J3})
we then obtain
\beqa 
\label{J23final} 
J_1(p) & = & \frac{i}{|e|B}\sum_{\ell = 0}^\infty
\frac{D^\prime_\ell(a_p)}{\psi_p - 2\ell} \,, \nonumber\\
J_2(p) & = & \frac{i}{|e|B}\sum_{\ell = 0}^\infty
\frac{D^{\prime\prime}_\ell(a_p)} {\psi_p - 2\ell} \,, 
\eeqa
where
\beqa 
D^\prime_n(a) & = & \frac{dD_n(a)}{da} \,, \nonumber\\
D^{\prime\prime}_n(a) & = & \frac{d^2D_n(a)}{da^2} \,. 
\eeqa
Furthermore, from Eq.\ (\ref{Ki}), it follows that
\beqa \label{Kifinal} K_0(p) & = & 2\pi\sum_{n = 0}^\infty
D_n(a_p)\delta(p^2_\parallel - m^2_e - 2n|e|B)\,, \nonumber\\
K_1(p) & = & 2\pi\sum_{n = 0}^\infty
D^\prime_n(a_p)\delta(p^2_\parallel - m^2_e - 2n|e|B)\,,
\nonumber\\ K_2(p) & = & 2\pi\sum_{n = 0}^\infty
D^{\prime\prime}_n(a_p) \delta(p^2_\parallel - m^2_e - 2n|e|B) \,.
\eeqa

Substituting Eq.\ (\ref{intLsaexp}) into (\ref{pimunuefffinal}),
and using Eq.\ (\ref{Kifinal}) we then obtain
\beqa 
\label{pimunuefffinal2} 
\pi^{(eff)}_{\mu\nu}(\omega,\vec Q)
& = & 4ie^2\sum_{n = 0}^\infty\int\frac{d^3p}{(2\pi)^3 2E_n}
\left\{\left[ M^{(n)}_{\mu\nu}(p,q)[f_e(E_n) + f_{\bar e}(E_n)] +
(q\rightarrow -q)\right]\right.\nonumber\\ &&\mbox{} +
\left.\left[N^{(n)}_{\mu\nu}(p,q) [f_e(E_n) - f_{\bar e}(E_n)] -
(q\rightarrow -q)\right]\right\} \,. 
\eeqa
where
\beq 
p^0 = E_n \,,
\eeq
with
\beq 
E_n = \sqrt{\vec p_\parallel^{\;2} + m^2_e + 2n|e|B} \,, 
\eeq
and
\beqa 
\label{MN} M^{(n)}_{\mu\nu}(p,q) & = & -2m^2_e
D^\prime_n(a_p)J_1(p + q) R_{\mu\nu} + m^2_e\left[D_n(a_p)J_0(p +
q) + D^\prime_n(a_p)J_1(p + q)\right]g_{\mu\nu}\nonumber\\
&&\mbox{} - g_{\mu\nu}\left[D_n(a_p)J_0(p + q) -
D^\prime_n(a_p)J_1(p + q)\right](p^2 + p\cdot q) \nonumber\\ & &
\mbox{} - g_{\mu\nu}\left[ -D^{\prime\prime}_n(a_p)J_0(p + q) -
D_n(a_p)J_2(p + q)\right.\nonumber\\ &&\mbox{} \left. +
D^\prime_n(a_p)J_1(p + q) + D^{\prime\prime}_n(a_p)J_2(p + q)
\right](p^2_\perp + p_\perp\cdot q_\perp) \nonumber\\ & & \mbox{}
+ \left[g_{\mu\alpha}g_{\nu\beta}D_n(a_p)J_0(p + q) -
R_{\mu\alpha}g_{\nu\beta}D^{\prime\prime}_n(a_p)J_0(p +
q)\right.\nonumber\\ &&\mbox{} \left. -
g_{\mu\alpha}R_{\nu\beta}D_n(a_p)J_2(p + q) +
R_{\mu\alpha}R_{\nu\beta}D^{\prime\prime}_n(a_p)J_2(p +
q)\right.\nonumber\\ &&\mbox{} +
\left.U_{\mu\alpha}U_{\nu\beta}D^\prime_n(a_p)J_1(p + q)  +
(\mu\leftrightarrow\nu) \right]p^\alpha(p + q)^\beta \,, \nonumber\\
N^{(n)}_{\mu\nu}(p,q) & = & i\epsilon_{\mu\nu\alpha\beta}\left[
D^\prime_n(a_p)J_0(p + q)g^{\alpha\lambda}U^{\beta\rho} +
D_n(a_p)J_1(p + q)U^{\alpha\lambda}g^{\beta\rho}\right.\nonumber\\
&&\mbox{} - \left.D^\prime_n(a_p)J_2(p +
q)R^{\alpha\lambda}U^{\beta\rho} - D^{\prime\prime}_n(a_p)J_1(p +
q)U^{\alpha\lambda}R^{\beta\rho}\right] p_\lambda(p +
q)_\rho\nonumber\\ &&\mbox{} + m^2_e
i\epsilon_{\mu\nu\alpha\beta}u^\alpha b^\beta
\left[D^\prime_n(a_p)J_0(p + q) - D_n(a_p)J_1(p + q)\right] \,.
\eeqa
Using Eq.\ (\ref{pimunuefffinal2}) as the starting point, 
the form factors that enter in the dispersion relations
of the propagating modes can be computed, as we show next.

%
% sec4
%
\section{Dispersion relations of the normal modes}
\label{sec:disprel}

\subsection{General case}
\label{subsec:general}
As already argued in Section\ \ref{sec:kinematics}, the photon
self-energy can be expressed as shown in Eq.\ (\ref{gendecomp}).
The nine independent coefficients $\pi^{(ij)}(\omega,Q)$ 
are then determined to one-loop order by
\beq 
\label{defpielements} 
\pi^{(ij)}(\omega,\vec{Q}) =
-\epsilon^\mu_i(\vec Q)\epsilon^\nu_j(\vec Q)
\pi^{(eff)}_{\mu\nu}(\omega,\vec{Q})\,, 
\eeq
with $\pi^{(eff)}_{\mu\nu}(\omega,\vec{Q})$ given by
Eq.\ (\ref{pimunuefffinal2}). While $\epsilon^\mu_{1}(\vec Q)$ and
$\epsilon^\mu_{2}(\vec Q)$ are independent of $Q$,
$\epsilon^\mu_{3}(\vec Q)$ is not. We will denote by
$\epsilon^\mu_{i}(0)$ the basis vectors for $Q = 0$, and therefore
\beqa
\label{defepsilonsQ0}
\epsilon^\mu_{1,2}(0) & = & \epsilon^\mu_{1,2}(\vec Q) \,, \nonumber\\
\epsilon^\mu_{3}(0) & = & (0,\hat e_3) \,.
\eeqa
It is useful to note that
\beq
\label{epsilon3Q}
\epsilon^\mu_{3}(\vec Q) = \frac{Q}{\sqrt{q^2}}u^\mu +
\frac{\omega}{\sqrt{q^2}}\epsilon^\mu_{3}(0) \,,
\eeq
and by using
\beq
q^\mu = \omega u^\mu + Q\epsilon^\mu_3(0) \,,
\eeq
we can also write
\beq
\label{epsilon3Q2}
\epsilon^\mu_3(\vec Q) = \frac{Q}{\sqrt{q^2}}q^\mu +
\frac{\sqrt{q^2}}{\omega}\epsilon^\mu_3(0) \,.
\eeq

The polarization vectors of the various propagating modes
can be expanded in the form
\beq
\label{xiexpansion}
\xi^\mu = \sum_{i = 1,3}\alpha_i(\vec Q)\epsilon^\mu_i(\vec Q) \,,
\eeq
where the coefficients $\alpha_i(\vec Q)$, and the corresponding
dispersion relations, are determined by solving the equation
\beq
\label{fieldeq}
(q^2\tilde g_{\mu\nu} - \pi_{\mu\nu})\xi^\nu = 0\,.
\eeq
In matrix notation, Eq.\ (\ref{fieldeq}) can be written in the
form
\beq 
\label{matrixeq} (q^2 - \Pi){\bf\alpha} = 0 \,, 
\eeq
where
\beq
{\bf\alpha} = \left(\begin{array}{c}\alpha_1\\ \alpha_2\\
\alpha_3\end{array}\right) \,, 
\eeq
and $\Pi$ is the matrix formed by the coefficients $\pi^{(ij)}$.
The formulas for the functions $\pi^{(ij)}$ are
obtained by applying Eq.\ (\ref{defpielements}).
For that purpose, it is useful to note that
we can make the replacement
\beq
\epsilon^\mu_3(\vec Q) \rightarrow 
\frac{\sqrt{q^2}}{\omega}\epsilon^\mu_3(0)\,,
\eeq
which follows by using Eq.\ (\ref{epsilon3Q2}) and the transversality
condition satisfied by $\pi^{(eff)}_{\mu\nu}$. Thus we obtain
\beqa \label{piijformulas} \pi^{(ab)}(\omega,\vec{Q}) & =
&-\pi^{(eff)\,ab}(\omega,\vec{Q})  \nonumber\\
\pi^{(a3)}(\omega,\vec{Q}) & = & -\frac{\sqrt{q^2}}{\omega}
\pi^{(eff)\,a3}(\omega,\vec{Q}) \nonumber\\
\pi^{(3a)}(\omega,\vec{Q}) & = & -\frac{\sqrt{q^2}}{\omega}
\pi^{(eff)\,3a}(\omega,\vec{Q}) \nonumber\\
\pi^{(33)}(\omega,\vec{Q}) & = &
-\frac{q^2}{\omega^2}\pi^{(eff)\,33}(\omega,\vec{Q}) \,, \eeqa
where the indices $a,b$ can take the values between $1$ and $2$.
While Eq.\ (\ref{piijformulas}) allows us to read the elements
$\pi^{(ij)}$ off Eq.\ (\ref{pimunuefffinal}) by inspection,
the problem of finding the dispersion relations in the
general case is a formidable one.
For this reason we now consider in some detail the so called
long wavelength limit, which is a particularly important case
that holds in a variety of physical applications.

The point to stress here is that the nine elements $\pi^{(ij)}$
determined by Eq.\ (\ref{piijformulas}), form a complete set
of independent functions that parametrize the photon self-energy
in the most general way that is consistent with the transversality condition.
In particular, this characteristic remains valid independently
of any approximation that may be used to compute the elements
$\pi^{(eff)\,ij}$ that must be substituted in the right-hand side
in Eq.\ (\ref{piijformulas}).

\subsection{Long wavelength limit}
\label{subsec:lwlimit}
The long wavelength ($\limQzero$) limit is particularly important
and we consider it separately  as a special case. We stress that,
while we are considering this particular limiting case,
we do not make any assumptions about the conditions of the gas
or the magnitude of the magnetic field. Therefore, the formulas
that we obtain below hold for relativistic or non-relativistic
gases, whether they are degenerate or not, and for any value
of the magnetic field. This limit is valid under the condition
\beq 
\label{longwavelength} \omega \gg  v_e Q \,, 
\eeq
where $v_e$ stands for the average velocity of an electron in the
gas. In the limit $\limQzero$, we can write
\beq 
q^\mu = \omega u^\mu \,,
\eeq
and therefore only $u^\mu$ and $b^\mu$ are independent vectors.
The most general form of $\pi^{(eff)}_{\mu\nu}$ consistent with
the transversality condition in this case is then
\beq 
\label{pimunuQzero} 
\pi^{(eff)}_{\mu\nu}(\omega,\limQzero) =
\pi_T(\omega,\limQzero) R_{\mu\nu} + \pi_L(\omega,\limQzero)
Q_{\mu\nu} + \pi_P(\omega,\limQzero)P_{\mu\nu} \,, 
\eeq
where $R_{\mu\nu}, P_{\mu\nu}$ and $Q_{\mu\nu}$ are defined in 
Eq.\ (\ref{deftensors}),
and  the functions $\pi_{T,L,P}$ are determined from the one-loop
expression for $\pi_{\mu\nu}^{(eff)}$ by means of the projection
formulas
\beqa 
\label{projections} \pi_T(\omega,\vec{Q}) & = &
\frac{1}{2}R^{\mu\nu}\pi_{\mu\nu}^{(eff)}(\omega,\vec{Q})\,, \nonumber\\ 
\pi_L(\omega,\vec{Q}) & = &
Q^{\mu\nu}\pi_{\mu\nu}^{(eff)}(\omega, \vec{Q}) \,, \nonumber\\
\pi_P(\omega,\vec{Q}) & = &
-\frac{1}{2}P^{\mu\nu}\pi_{\mu\nu}^{(eff)}(\omega,\vec{Q}) \,.
\eeqa
Although we are not indicating it explicitly, the functions
$\pi_{T,L,P}(\omega,\vec Q)$ depend, in addition, on the magnetic
field $B$. Substituting Eq.\ (\ref{pimunuefffinal}) into
(\ref{projections}) we obtain
\beqa 
\label{piTLP} \pi_{T,L}(\omega,\vec{Q}) & = & 
4ie^2\sum_{n = 0}^\infty\int\frac{d^3p}{(2\pi)^3 2E_n} 
[f_e(E_n) + f_{\bar e}(E_n)]\left[I^{(n)}_{T,L}(p,q) + 
(q\rightarrow -q)\right] \,, \nonumber\\[12pt]
\pi_P(\omega,\vec{Q}) & = & 
4ie^2\sum_{n =0}^\infty\int\frac{d^3p}{(2\pi)^3 2E_n} 
[f_e(E_n) - f_{\bar e}(E_n)]\left[I^{(n)}_P(p,q) - 
(q\rightarrow -q)\right] \,, 
\eeqa
where
\beqa 
\label{contractions} 
I^{(n)}_T & \equiv &
\frac{1}{2}R^{\mu\nu}M^{(n)}_{\mu\nu}\,,\nonumber\\[12pt] 
I^{(n)}_L & \equiv & Q^{\mu\nu}M^{(n)}_{\mu\nu} \,,\nonumber\\[12pt] 
I^{(n)}_P & \equiv & -\frac{1}{2}P^{\mu\nu}N^{(n)}_{\mu\nu} \,. 
\eeqa
By direct calculation, it is straightforward to obtain, for any
$q^\mu$, the following expressions
\beqa 
\label{ITLP} 
I^{(n)}_T & =  & [D_n(a_p)J_0(p + q) - D^\prime_n(a_p)J_1(p + q)] 
h_T(p,q)\,, \nonumber\\[12pt] 
I^{(n)}_L  & = & -[D_n(a_p)J_0(p + q) + D^{\prime\prime}_n(a_p)J_2(p + q) -
D_n(a_p)J_2(p + q) - D^{\prime\prime}_n(a_p)J_0(p + q)]h_{L1}(p,q)
\nonumber\\ 
&&\mbox{} - [D_n(a_p)J_0(p + q) + D^\prime_n(a_p)J_1(p+ q)] h_{L2}(p,q)\,,
\nonumber\\[12pt] 
I^{(n)}_P & = & [D_n(a_p)J_1(p + q) - D^\prime_n(a_p)J_0(p + q)]h_T(p,q)\,, 
\eeqa
where
\beqa 
\label{hTL} 
h_T & =  & (p\cdot b)^2 - (p\cdot u)^2 + (p\cdot b)(q\cdot b) - 
(p\cdot u)(q\cdot u) + m^2_e \,, \nonumber\\[12pt]
h_{L1} & = & p^2_\perp +  p_\perp\cdot q_\perp \,, \nonumber\\[12pt]
h_{L2} & = & (p\cdot u)^2 + (p\cdot b)^2 + (p\cdot b)(q\cdot b) +
(p\cdot u)(q\cdot u) - m^2_e \,. 
\eeqa
In order to evaluate the functions $\pi_{T,L,P}$ in the
$\limQzero$ limit, as indicated in Eq.\ (\ref{pimunuQzero}), we
proceed as follows. Remembering that $p^0 = E_n$, in this limit
\beq
\psi_{p + q} = \frac{1}{|e|B}(\omega^2 + 2\omega E_n + 2n|e|B) \,,
\eeq
and then from Eq.\ (\ref{J1final})
\beq 
J_0(p + q) = 
i\sum_\ell \frac{D_\ell(a_p)}{\omega^2 + 2\omega
E_n + 2(n - \ell)|e|B)} \,, 
\eeq
with similar formulas for $J_{1,2}(p + q)$, but with $D_\ell(a_p)$
being replaced by $D^\prime_\ell(a_p)$ and
$D^{\prime\prime}_\ell(a_p)$, respectively. Taking the $\limQzero$
limit  of the expressions in Eq.\ (\ref{hTL}) and substituting
them in Eqs.\ (\ref{ITLP}) we then obtain
\beqa 
\label{ITLPQ0} 
\left.I^{(n)}_T\right|_{\limQzero} & = &
-i\sum_\ell \frac{\left(D_n(a_p)D_\ell(a_p) - 
D^\prime_n(a_p)D^\prime_\ell(a_p)\right)\left( \omega E_n + 2n|e|B\right)} 
{\omega^2 + 2\omega E_n + 2(n - \ell)|e|B}\,,\nonumber\\ 
\left.I^{(n)}_L\right|_{\limQzero} & = &
ia_p|e|B\sum_\ell \frac{D_n(a_p)D_\ell(a_p) +
D^{\prime\prime}_n(a_p)D^{\prime\prime}_\ell(a_p) -
D_n(a_p)D^{\prime\prime}_\ell(a_p) - D^{\prime\prime}_n(a_p)D_\ell(a_p)} 
{\omega^2 + 2\omega E_n + 2(n - \ell)|e|B} \nonumber\\ &&\mbox{} -
i\sum_\ell\frac{\left[D_n(a_p)D_\ell(a_p) +
D^\prime_n(a_p)D^\prime_\ell(a_p)\right] \left[2E^2_n + \omega E_n
- 2(m^2_e + n|e|B)\right]} {\omega^2 + 
2\omega E_n + 2(n - \ell)|e|B} \,,\nonumber\\ 
\left.I^{(n)}_P\right|_{\limQzero} & = &
-i\sum_\ell\frac{\left(D_n(a_p)D^\prime_\ell(a_p) -
D^\prime_n(a_p)D_\ell(a_p)\right)\left(\omega E_n + 2n|e|B\right)}
{\omega^2 + 2\omega E_n + 2(n - \ell)|e|B} \,. 
\eeqa
With the help of the identities,
\beqa 
\frac{d}{dx}\left(L_{n} - L_{n-1}\right) & = & -L_{n-1}\,, \nonumber\\ 
x\frac{dL_n}{dx} & = & n(L_n - L_{n-1}) \,, 
\eeqa
the following is easily derived
\beq 
a\left(D_n(a) - D^{\prime\prime}_n(a)\right) = 2nD_n(a) \,,
\eeq
so that the formula for $I^{(n)}_L|_{\limQzero}$ becomes
\beqa 
\left.I^{(n)}_L\right|_{\limQzero} & = &
i\frac{|e|B}{a_p}\sum_\ell \frac{4n\ell D_n(a_p)D_\ell(a_p)}
{\omega^2 + 2\omega E_n + 2(n - \ell)|e|B} \nonumber\\ 
&&\mbox{} -
i\sum_\ell\frac{\left[D_n(a_p)D_\ell(a_p) +
D^\prime_n(a_p)D^\prime_\ell(a_p)\right] \left[2E^2_n + \omega E_n
- 2(m^2_e + n|e|B)\right]} {\omega^2 + 2\omega E_n + 2(n - \ell)|e|B} \,. 
\eeqa

The integrand in Eq.\ (\ref{piTLP}) depends on $\vec p_{\perp}$
only through $a_p$, so that we can replace $d^3p \rightarrow
dp_{\parallel}(\pi|e|Bda_p)$. With the help of the normalization
condition satisfied by the Laguerre polynomials
\beqa 
\int_0^\infty dx\,e^{-x}L_m(x)L_n(x) & = & \delta_{n,m} \,,
\eeqa
where $\delta_{n,m}$ is the Kronecker symbol with the convention
\beq 
\delta_{n,n} = 0 \qquad \mbox{for $n < 0$} \,, 
\eeq
the following hold
\beqa 
\int_0^\infty da D_n(a)D_m(a) & = & \frac{(-1)^{n + m}}{2}
\left[\delta_{n,m} - \delta_{n-1,m} - \delta_{n,m-1} + 
\delta_{n-1,m-1}\right] \,,\nonumber\\
\int_0^\infty da D^\prime_n(a)D^\prime_m(a) & = & \frac{(-1)^{n +
m}}{2} \left[\delta_{n,m} + \delta_{n-1,m} + \delta_{n,m-1} +
\delta_{n-1,m-1}\right] \,,\nonumber\\
\int_0^\infty da D_n(a)D^\prime_m(a) & = & -\frac{(-1)^{n + m}}{2}
\left[\delta_{n,m} - \delta_{n-1,m} + \delta_{n,m-1} -
\delta_{n-1,m-1}\right] \,,\nonumber\\ 
\int_0^\infty \frac{da}{a} D_n(a)D_m(a) & = & \frac{1}{n}\delta_{n,m} \,. 
\eeqa
Thus, defining
\beq 
\pi_X(\omega) \equiv \pi_X(\omega,\limQzero) \qquad
\mbox{(for $X = T,L,P$)} \,,
\eeq
by substituting Eq.\ (\ref{ITLPQ0}) in Eq.\ (\ref{piTLP}) and
using the above integration formulas we obtain
\beq
\label{piTLPgeneral} 
\pi_X(\omega) = \frac{|e|^3B}{4\pi^2}\sum_{n = 0}\pi_X^{(n)}(\omega) \,,
\eeq
where
\beqa \label{piTLPn}
\pi^{(0)}_T{(\omega)} & = & \int_{-\infty}^\infty dp_{\parallel}
(f_e(E_0) + f_{\bar e}(E_0))\frac{\omega}{\omega^2 + 2\omega E_0 -
2|e|B} + (\omega \rightarrow -\omega) \,,\nonumber\\
\pi^{(n\not= 0)}_T{(\omega)} & = & \int_{-\infty}^\infty
\frac{dp_{\parallel}}{E_n} (f_e(E_n) + f_{\bar
e}(E_n))\left(\omega E_n + 2n|e|B\right) \left[\frac{1}{\omega^2 +
2\omega E_n - 2|e|B} + \frac{1}{\omega^2 + 2\omega E_n +
2|e|B}\right] + (\omega \rightarrow -\omega) \,, \nonumber\\
\pi^{(0)}_P{(\omega)} & = & \int_{-\infty}^\infty dp_{\parallel}
(f_e(E_0) - f_{\bar e}(E_0))\frac{\omega}{\omega^2 + 2\omega E_0 -
2|e|B} - (\omega \rightarrow -\omega) \,,\nonumber\\
\pi^{(n\not= 0)}_P{(\omega)} & = & \int_{-\infty}^\infty
\frac{dp_{\parallel}}{E_n} (f_e(E_n) - f_{\bar
e}(E_n))\left(\omega E_n + 2n|e|B\right) \left[\frac{1}{\omega^2 +
2\omega E_n - 2|e|B} - \frac{1}{\omega^2 + 2\omega E_n +
2|e|B}\right] - (\omega \rightarrow -\omega) \,, \nonumber\\
\pi^{(0)}_L{(\omega)} & = & -\int_{-\infty}^\infty
\frac{dp_{\parallel}}{E_0} (f_e(E_0) + f_{\bar
e}(E_0))\frac{2m^2_e}{\omega^2 + 2\omega E_0} + 
(\omega\rightarrow -\omega)\,,\nonumber\\
\pi^{(n\not= 0)}_L{(\omega)} & = & -\int_{-\infty}^\infty
\frac{dp_{\parallel}}{E_n} (f_e(E_n) + f_{\bar
e}(E_n))\frac{8n|e|B + 4m^2_e}{\omega^2 + 2\omega E_n} +
(\omega\rightarrow -\omega) \,. 
\eeqa
Since we are concerned only with the real part of the self-energy
in this work, the above integrals are actually defined in the
sense of their principal value part. 

As we discuss next, the
formulas in Eq.\ (\ref{piTLPn}) allow us to study the dispersion
relations in the long wavelength limit.
In this case  Eq.\ (\ref{matrixeq}) becomes
\beq
\label{dispiQ0}
[q^2 - \Pi_0]\tilde\alpha = 0 \,. \eeq
Here the matrix $\Pi_0$ is formed by the coefficients $\pi^{(ij)}$
that are determined from Eq.\ (\ref{piijformulas}), but using the
formula for $\pi^{(eff)}_{\mu\nu}(\omega,\limQzero)$ that is given
by Eq.\ (\ref{pimunuQzero}).  Thus, 
using the kinematical relations given in Sec.\ \ref{sec:kinematics},
by straightforward algebra we find
\beq \label{pitildeQ0} \Pi_0 = \left(\begin{array}{ccc}
\pi_T(\omega)\cos^2\theta + \pi_L(\omega)\sin^2\theta &
-i\pi_P(\omega)\cos\theta &
\frac{\sqrt{q^2}}{\omega}[\pi_L(\omega) -
\pi_T(\omega)]\sin\theta\cos\theta
\\
i\pi_P(\omega)\cos\theta & \pi_T(\omega) &
-\frac{\sqrt{q^2}}{\omega}i\pi_P(\omega)\sin\theta
\\
\frac{\sqrt{q^2}}{\omega}[\pi_L(\omega) -
\pi_T(\omega)]\sin\theta\cos\theta &
\frac{\sqrt{q^2}}{\omega}i\pi_P(\omega)\sin\theta &
\frac{q^2}{\omega^2}\left[\pi_T(\omega)\sin^2\theta +
\pi_L(\omega)\cos^2\theta\right]
\end{array}\right) \,.
\eeq
It should be remembered that in Eq. (\ref{dispiQ0}), $\tilde\alpha$
depends on $\vec Q$. In what follows, we consider some particular
cases.

\subsubsection{$Q = 0$}
We consider first the case the zero momentum limit of the
dispersion relations, which corresponds to a photon traveling
with a vanishingly small momentum in an arbitrary direction. The
matrix $\Pi_0$ in this case is given as in Eq.\ (\ref{pitildeQ0})
but with the replacement $q^2/\omega^2\rightarrow 1$. The
eigenvectors of the matrix in this ($Q = 0$) case are found to be
given by
\beqa 
\tilde\alpha_{\pm} & = & \left(\begin{array}{c} \cos\theta
\\ \pm i \\ -\sin\theta
\end{array}\right) \,,\nonumber\\[12pt]
\tilde\alpha_{L} & = & \left(\begin{array}{c} \sin\theta \\ 0 \\
\cos\theta
\end{array}\right) \,,
\eeqa
with corresponding eigenvalues $\pi^0_T \pm \pi^0_P$ and
$\pi^0_L$, respectively.  Therefore, from Eq.\
(\ref{xiexpansion}), in the $Q \rightarrow 0$ limit the
polarization vectors of the normal modes are given by
\beqa 
\label{polvectorsQ0} 
\xi^\mu_{\pm}(0) & = &
\cos\theta\epsilon^\mu_1(\vec 0) \pm i\epsilon^\mu_2(\vec 0) -
\sin\theta\epsilon^\mu_3(\vec 0) \,,\nonumber\\
\xi^\mu_{L}(0) & = & \sin\theta\epsilon^\mu_1(\vec 0) +
\cos\theta\epsilon^\mu_3(\vec 0) \,, 
\eeqa
and the corresponding dispersion relations are determined by
solving
\beqa 
\label{disprelQ0} 
\omega^2 - \left[\pi_T \pm \pi_P \right] & = & 0 \,,\nonumber\\ 
\omega^2 - \pi_L & = & 0 \,, 
\eeqa
respectively.

\subsubsection{Propagation parallel to the magnetic field}
This corresponds to set
\beq 
\theta = 0 
\eeq
in Eq.\ (\ref{pitildeQ0}). However, in contrast to the previous
case, we do not set $Q = 0$ (although we still assume that $Qv_e
\ll \omega$, which is the condition under which Eq.\
(\ref{pitildeQ0}) holds). In this case,
\beq 
\label{pitildeQ0parallel} 
\Pi_0(\omega,\vec Q) =
\left(\begin{array}{ccc} \pi^0_T & -i\pi^0_P & 0 \\
i\pi^0_P & \pi_T & 0 \\
0 & 0 & \frac{q^2}{\omega^2}\pi^0_L
\end{array}\right) \,,
\eeq
and therefore
\beq 
\tilde\alpha_{\pm}  =  \left(\begin{array}{c} 1 \\ \pm i
\\ 0
\end{array}\right)\,,
\qquad
\tilde\alpha_{L}  =  \left(\begin{array}{c} 0 \\ 0 \\ 1
\end{array}\right) \,,
\eeq
with corresponding eigenvalues $\pi^0_T \pm \pi^0_P$ and
$\frac{q^2}{\omega^2}\pi^0_L$, respectively. The polarization
vectors of the normal modes are given by
\beqa 
\label{polvectorsQ0parallel} 
\xi^\mu_{\pm}(\vec Q) & = &
\frac{1}{\sqrt{2}}\left( \epsilon^\mu_1(\vec Q) \pm 
i\epsilon^\mu_2(\vec Q)\right) \,,\nonumber\\
\xi^\mu_{L}(\vec Q) & = & \epsilon^\mu_3(\vec Q) \,,
\eeqa
and the corresponding dispersion relations are determined by
solving
\beqa 
\label{disprelQ0parallel} \omega^2 - Q^2 - 
\left[\pi_T \pm \pi_P \right] & = & 0 \,, \nonumber\\ 
\omega^2 - \pi_L & = & 0 \,,
\eeqa
respectively. One of the characteristics of this solution is the
fact that the longitudinal mode dispersion relation is independent
of $Q$, which is a well known feature of this case.

\subsubsection{Propagation perpendicular to the magnetic field}
In this case
\beq 
\theta = \pi/2 
\eeq
and, as in the previous one, we maintain $Q \not = 0$. 
Then, from Eq.\ (\ref{pitildeQ0}),
\beq 
\label{pitildeQ0perp} 
\Pi_0(\omega,\vec Q) =
\left(\begin{array}{ccc} \pi_L & 0 & 0 \\
0 & \pi_T & -i\frac{\sqrt{q^2}}{\omega}\pi_p \\
0 & i\frac{\sqrt{q^2}}{\omega}\pi_p &
\frac{q^2}{\omega^2}\pi_T
\end{array}\right) \,.
\eeq
The dispersion relation of the longitudinal mode [with the
polarization vector $\xi^\mu_L(\vec Q) = \epsilon^\mu_1(\vec Q)$]
is obtained by solving
\beq 
\omega^2 - Q^2 - \pi_L  = 0 \,.
\eeq
For the transverse modes it becomes
\beq (\omega^2 - \pi_T)^2 - Q^2(\omega^2 - \pi_T) - \pi^{2}_P = 0 \,, 
\eeq
which can be written in the alternate form
\beq 
\frac{Q^2}{\omega^2} = \frac{\left(1 - \frac{\pi_T}{\omega^2}\right)^2 -
\left(\frac{\pi_P}{\omega^2}\right)^2} {1 - \frac{\pi_T}{\omega^2}} \,.
\eeq
The corresponding eigenvectors are linear combinations of 
$\epsilon^\mu_{2,3}(\vec Q)$.

For an arbitrary direction of propagation, the longitudinal and
transverse polarizations mix, so that the propagating modes are
neither purely longitudinal nor purely transverse. In this case,
a perturbative solution, that is valid for
sufficiently small $Q$, can obtained.

%
% sec5
%
\section{Explicit formulas}
\label{sec:explicitformulas}

Although the formulas in Eq.\ (\ref{piTLPn}) have been obtained by
taking the long wavelength ($Q\rightarrow 0$) limit,
no approximations have been made. However, they
can be calculated explicitly
for a variety of situations, under certain approximations. Here
we specifically consider the case in which $\omega$ satisfies
\begin{equation}
\label{smallomega}
\omega \ll \sqrt{|e|B}
\end{equation}
and
\beq
\label{smallwcondL}
\omega \ll 2\langle E_e\rangle \,,
\eeq
where $\langle E_e\rangle$ stands for an average value of the energy of
an electron in the gas.
Thus, if the gas is non-relativistic,
the condition holds for $\omega \ll m_e$. If the gas is
extremely relativistic, it  also holds
for $\omega > m_e$, subject to the condition in Eq.\ (\ref{smallwcondL}).
Together with the condition given in
Eq.\ (\ref{longwavelength}), this implies that we are considering a regime
in which
\beq
\label{longwavelengthcond}
v_e Q \ll \omega \ll \sqrt{|e|B}\,,\; 2\langle E_e\rangle \,,
\eeq
to which we will refer as the \emph{low frequency regime}.

Then, under the assumption that the condition in Eq.\ (\ref{smallomega}) holds,
the formulas for $\pi^{(n)}_{T,P}$ given in Eq.\ (\ref{piTLPn}) reduce to
\beqa
\label{piTPnsmallw}
\pi^{(n)}_T{(\omega)}
& = & (2 - \delta_{n,0})\omega^2\int_{-\infty}^\infty dp_{\parallel}
(f_e(E_n) + f_{\bar e}(E_n))\frac{1}{\omega^2 E^2_n - |e|^2 B^2}
\left[E_n - \frac{n|e|B}{E_n}\right] \,,\nonumber\\[12pt]
\pi^{(n)}_P{(\omega)} & = & (2 - \delta_{n,0})
(\omega |e|B)\int_{-\infty}^\infty dp_{\parallel}
(f_e(E_n) - f_{\bar e}(E_n))\frac{1}{\omega^2 E^2_n - |e|^2 B^2}
\left[1 - \frac{2n|e|B}{E^2_n}\right] \,,
\eeqa
and similarly,
\beq
\label{piLnsmallw}
\pi^{(n)}_L(\omega)  = (2 - \delta_{n,0})
\int_{-\infty}^\infty dp_{\parallel} (f_e(E_n) + f_{\bar e}(E_n))
\frac{(m_e^2 + 2n|e|B)}{E^3_n}
\eeq
assuming Eq.\ (\ref{smallwcondL}).
We should note here that the judicious application of the
low frequency regime conditions to Eq.\ (\ref{piTLPn}) requires
some care. The procedure we have followed to arrive at
Eqs.\ (\ref{piTPnsmallw}) and (\ref{piLnsmallw}) is outlined
in Appendix\ \ref{app:lowfrequencyregime}.

We stress once more that these formulas hold for any conditions
of the electron gas, whether it is relativistic or non-relativistic,
and degenerate or not, and for any value of the magnetic field,
subject to Eq.\ (\ref{smallomega}).
However, for specific conditions of the gas,
these expressions reduce further to simpler formulas.
We consider several examples separately.

\subsection{Nonrelativistic gas}
\label{subsec:NRlimit}

In this case the temperature and chemical potential
of the electron gas are assumed to be such that
\beq
\label{nrconditions}
\beta m_e \gg 1 \,, \qquad \frac{m_e}{\mu} \gg 1\,.
\eeq
This implies that
\beq
f_{\bar e} \approx 0\,,
\eeq
while for $f_e$, only values
those values of $p_\parallel$ and $n$ for which
\beq
E_n \simeq m_e
\eeq
contribute significantly in the integrals
because otherwise $f_e$ becomes small.
Therefore, to the lowest order in $1/m_e$ we can replace
\beq
E_n \rightarrow m_e
\eeq
in the integrands in Eqs.\ (\ref{piTPnsmallw}) and
(\ref{piLnsmallw}), and retaining the $O(1/m_e)$ terms
leads to
\beqa
\label{piTLPsmallwnr}
\pi_T(\omega) & = &
\frac{\omega^2\Omega_0^2}{\omega^2 - \omega_B^2} \,,\nonumber\\
\pi_L(\omega) & = & \Omega_0^2 \,,\nonumber\\
\pi_P(\omega) & = & \frac{\omega\omega_B\Omega_0^2}{\omega^2 - \omega_B^2} \,,
\eeqa
where
\beqa
\label{omegaB}
\omega_B & \equiv & \frac{|e|B}{m_e} \,, \\
\label{omegap}
\Omega_0^2 & \equiv & \frac{e^2 n_e}{m_e} \,,
\eeqa
with
\beqa
\label{ne}
n_e & = &\frac{|e|B}{4\pi^2}
\left[\sum_{n = 0}^\infty(2 - \delta_{n,0})
\int_{-\infty}^\infty dp_{\parallel} f_e(E_n)\right] \,.
\eeqa
The formulas in Eqs.\ (\ref{omegaB}), (\ref{omegap}) and (\ref{ne})
are the standard expressions for the cyclotron frequency, the plasma
frequency and total electron number density, respectively. Moreover,
using the results of Section\ \ref{subsec:lwlimit},
the formulas in
Eq.\ (\ref{piTLPsmallwnr}) allows us to reproduce very simply the classic
results\cite{llphyskin} for the photon dispersion relations in the case
under consideration. For example, if the photon propagates
parallel to the magnetic field, the dispersion relations given
by Eq.\ (\ref{disprelQ0parallel}) become
\beqa
\label{disprelexample}
&&\omega^2-Q^2-\frac{\omega\Omega^2_0}{(\omega-\omega_B)}=0,\nonumber\\
&&\omega^2-Q^2-\frac{\omega\Omega^2_0}{(\omega+\omega_B)}=0,\nonumber\\
&&\omega=\Omega_0.
\eeqa

The formulas in Eq.\ (\ref{piTLPsmallwnr}), and whence those
in Eq.\ (\ref{disprelexample}), represent
the leading term in powers
of $1/m_e$, and they neglect entirely the momentum-dependent terms.
The latter can give non-negligible corrections at
higher temperatures or chemical potential.
However, while the leading terms given above depend only on the
total number densities and not on the shape of the distribution
function, the same is not true with the momentum-dependent corrections.
Thus, for example, the corrections are different if we consider
the gas to be degenerate or classical.

In what follows we determine the corrections
specifically for the case of a classical gas. Thus we can put
\beq
\label{EnNR}
E_n \approx m_e + \frac{\vec p^{\;2}_\parallel}{2m_e} + n\omega_B,
\eeq
and for the distribution function
\beq
f_e \approx e^{-\beta (E_n - \mu)} \,.
\eeq
We consider first $\pi_{T,P}$. Using Eq.\ (\ref{EnNR})
we write the denominators of the integrands in Eq.\ (\ref{piTPnsmallw})
in the form
%
%\beq
%\omega^2 E^2_n - |e|^2 B^2 = m^2_e(\omega^2 - \omega^2_B)\left[
%1 + \frac{\omega^2}{\omega^2 - \omega^2_B}\left(
%\frac{p^2_\parallel}{m_e^2} + \frac{2n\omega_B}{m_e}\right)\right]
%\eeq
%
%and therefore
%
\beq
\frac{1}{\omega^2 E^2_n - |e|^2 B^2} \simeq
\frac{1}{m^2_e(\omega^2 - \omega^2_B)}\left[
1 - \frac{\omega^2}{\omega^2 - \omega^2_B}\left(
\frac{\vec p^{\;2}_\parallel}{m_e^2} + \frac{2n\omega_B}{m_e}\right)\right] \,,
\eeq
and similarly
\beqa
E_n - \frac{n|e|B}{E_n} & = &
m_e\left(1 + \frac{\vec p^{\;2}_{\parallel}}{m^2_e}\right)\,,\nonumber\\
1 - \frac{2n|e|B}{E^2_n} & = & 1 - \frac{2n\omega_B}{m_e} \,,
\eeqa
up to terms that are most linear in $\vec p^{\;2}_{\parallel}/m^2_e$ or
$\omega_B/m_e$. The integrals over $p_\parallel$ are carried out
very simply, and for the sums over $n$ we use the formulas
\begin{eqnarray}
\label{IS}
S_0 & = &\sum_{n = 0}^\infty e^{-\beta n\omega_B} = \frac{1}{1
- e^{-\beta\omega_B}}\,, \nonumber\\
S_1 & = & \sum_{n = 1}^\infty ne^{-\beta n\omega_B} =
\frac{e^{-\beta\omega_B}}{(1 - e^{-\beta\omega_B})^2} =
\frac{1}{4\sinh^2\left(\frac{\beta\omega_B}{2}\right)} \,.
\end{eqnarray}
Then eliminating the chemical potential in favor of $n_e$ by
means of  Eq.\ (\ref{ne}), which in the present case yields the
relation
\begin{eqnarray}
\label{neNR} n_e & = & e^{-\beta(m_e - \mu)}
\left(\frac{|e|B}{4\pi^2}\right) \sqrt{\frac{2m\pi}{\beta}}
(2S_0 - 1)\,,\nonumber\\
& = & e^{-\beta(m_e - \mu)}\left(\frac{|e|B}{4\pi^2}\right)
\sqrt{\frac{2m\pi}{\beta}}
\frac{1}{\tanh\left(\frac{\beta\omega_B}{2}\right)}\,,
\end{eqnarray}
this procedure yields
\beqa
\label{deltadef}
\pi_T(\omega) & = & \frac{\omega^2\Omega_0^2}{\omega^2 - \omega_B^2}
(1 + \delta_0) -
\frac{\omega^4\Omega_0^2}{(\omega^2 - \omega_B^2)^2}
(\delta_0 + \delta_1)\,,\nonumber\\
\pi_P(\omega) & = & \frac{\omega\omega_B\Omega_0^2}{\omega^2 - \omega_B^2}
(1 - \delta_1) - \frac{\omega^3\omega_B\Omega_0^2}{(\omega^2 - \omega_B^2)^2}
(\delta_0 + \delta_1) \,,
\eeqa
with
\begin{eqnarray}
\delta_0 & = & \frac{1}{\beta m_e} \,, \nonumber\\
\delta_1 & = & \frac{2\omega_B}{m_e\sinh\beta\omega_B} \,.
\end{eqnarray}
Notice that for $\beta\omega_B$ small, $\delta_1 \approx 2\delta_0$.
As $\beta\omega_B$ becomes larger, $\delta_1$ gives the
$O(\beta\omega^2_B/m_e)$ corrections to $\pi_T(\omega)$.

Turning the attention now to $\pi_L$, we
expand $E_n$ in the integrand of Eq.\ (\ref{piLnsmallw})
using Eq.\ (\ref{EnNR}). Retaining terms that are most linear
in $\vec p^{\;2}_\parallel/m_e^2$ or $\omega_B/m_e$ and proceeding as above
with the integrals and sums over $n$,
\beq
\label{piLnrclass}
\pi_L = \Omega_0^2\left[1 - \frac{3}{2m_e\beta} -
\frac{\omega_B}{m_e\sinh\beta\omega_B}\right]\,.
\eeq
In the limit $\beta\omega_B \rightarrow 0$, this formula reduces to
\beq
\pi_L = \Omega_0^2\left[1 - \frac{5}{2m_e\beta}\right]\,,
\eeq
which is the standard temperature correction to the
plasma frequency (in the non-relativistic, classical regime).
Otherwise, the last term in Eq.\ (\ref{piLnrclass})
gives the corrections of $O(\beta^2\omega^2_B/m_e)$
due to the presence of the magnetic field.

\subsection{Relativistic gas}
\label{subsec:ERlimit}
This limit corresponds to the conditions
\beq
1/\beta\,, \mu\,, \sqrt{|e|B} \gg m_e \,,
\eeq
so that we can effectively set
\beq
E_n \simeq \sqrt{\vec p^{\;2}_\parallel + 2|e|nB} \,,
\eeq
and we specifically consider the situation in which
\beq
\label{relcasesmallB}
1/\beta, \mu \gg \sqrt{|e|B} \,.
\eeq
Under these conditions
\beq
\label{smallB1}
\frac{1}{E_n}\frac{\delta E_n}{\delta n} \approx \frac{|e|B}{E^2_n} \ll 1 \,,
\eeq
(because $E_n$ is of order $1/\beta$ or $\mu$), and
we can then carry out the sums over $n$ by making the
replacement
\beq
\label{smallB2}
\sum_n \rightarrow \int dn = \int\frac{d^2\vec p_\perp}{2\pi|e|B} \,,
\eeq
where we have defined
\beq
\label{pperpaux}
\vec p^{\,2}_\perp \equiv 2n|e|B \,.
\eeq
With this substitution, the formula in Eq.\ (\ref{ne}) for the electron
number density, and the analogous one for the positron, reduce
to the standard expressions
\beq
\label{nestandard}
n_{e,\bar e} = 2\int \frac{d^3p}{(2\pi)^3} f_{e,\bar e}(E) \,,
\eeq
where, using Eq.\ (\ref{pperpaux}),
\beq
E = \sqrt{\vec p^{\;2}_\parallel + \vec p^{\,2}_\perp} = |\vec p|\,.
\eeq
In Eq.\ (\ref{nestandard}) we have
neglected the contribution from the $n = 0$ term, which is smaller
than the dominant terms by factors of
$\beta\sqrt{|e|B}$ and/or $\sqrt{|e|B}/\mu$.
Similarly, substituting
Eqs.\ (\ref{piTPnsmallw}) and (\ref{piLnsmallw}) into (\ref{piTLPgeneral}),
\beqa
\label{piTLPnsmallwrel}
\pi_T(\omega) & = & 2e^2\omega^2
\int\frac{d^3p}{(2\pi)^3} (f_e(E) + f_{\bar e}(E))
\frac{1}{\omega^2 E^2 - |e|^2 B^2}
\left[E - \frac{\vec p^{\,2}_\perp}{2E}\right] \,,\nonumber\\[12pt]
\pi_L(\omega) & = &
2e^2\int \frac{d^3p}{(2\pi)^3}
\frac{\vec p^{\,2}_\perp}{E^3}(f_e(E) + f_{\bar e}(E))\,,\nonumber\\[12pt]
\pi_P(\omega) & = &
2e^2\omega(|e|B)\int \frac{d^3p}{(2\pi)^3}
(f_e(E) - f_{\bar e}(E))
\frac{1}{\omega^2 E^2 - |e|^2 B^2}
\left[1 - \frac{\vec p^{\,2}_\perp}{E^2}\right] \,,
\eeqa
where we can replace $\vec p_\perp^{\;2}\rightarrow 2\vec p^{\;2}/3$
and $E\rightarrow |\vec p|$ in the integrand. The angular integration
is trivial but the remaining integration over $p$ cannot be carried out
exactly in general. We therefore
consider the classical and the degenerate limits, separately.

\subsubsection{Case I - Classical Gas}
In this case
\beq
\frac{1}{\beta} \gg \mu \,,
\eeq
and
\beq
f_{e,\bar e}(E) = e^{-\beta(E \mp \mu)} \,.
\eeq
The relationship between the chemical potential and the electron
charge density is then given by
\beq
n_e - n_{\bar e} = \frac{4}{\pi^2\beta^3}\sinh\beta\mu \simeq
\frac{4\mu}{\pi^2\beta^2} \,,
\eeq
and in addition
\beq
n_e + n_{\bar e} = \frac{4}{\pi^2\beta^3}\cosh\beta\mu \simeq
\frac{4}{\pi^2\beta^3} \,.
\eeq
The integration over $p$ is also trivial for $\pi_L$
and yields
\beq \pi_L = \frac{e^2\beta}{3}(n_e + n_{\bar e}) \,. \eeq
On the other hand, for $\pi_{T,P}$ the integration over
$p$ cannot be done exactly. The formulas for these quantities
can be expressed in the form
\beqa
\label{piTPerclass1} \pi_T & = & \frac{e^2\beta}{3}(n_e + n_{\bar e})
F_3\left(\frac{\Omega_{B}}{\omega}\right)\,,\nonumber\\
\pi_P & = & \frac{e^2\beta}{6}(n_e - n_{\bar e})
\frac{\Omega_B}{\omega}F_2\left(\frac{\Omega_B}{\omega}\right)\,,
\eeqa
where
\beq
\label{Fn}
F_n(z)  = \int_0^\infty dx\,e^{-x}\frac{x^n}{x^2 - z^2} \,,
\eeq
with
\beq
\Omega_B \equiv |e|B\beta \,.
\eeq
The quantity $\Omega_B$ plays in the present case the role
that $\omega_B$ occupies in the non-relativistic case. We can
obtain approximate values for the functions $F_n(z)$ by
considering the region in which $z \gg 1$ or $z \ll 1$.

The term $e^{-x}x^n$ is dominated by the values of $x \approx 1$.
For $z \gg 1$, the singularity at $x = z$ then lies outside the
effective region of integration, and we can substitute
\beq
\frac{1}{x^2 - z^2} \simeq \frac{1}{1 - z^2} \,.
\eeq
Therefore
\beq
F_n(z) \simeq \frac{n!}{1 - z^2} \,,
\eeq
and, from Eq.\ (\ref{piTPerclass1}), we can write
\beqa \pi_T & = & 2e^2\beta(n_e + n_{\bar e})
\left(\frac{\omega^2}{\omega^2 - \Omega^2_B}\right)\,,
\nonumber\\ \pi_P & = & \frac{e^2\beta}{3}(n_e - n_{\bar e})
\left(\frac{\omega\Omega_B}{\omega^2 -\Omega^2_B}\right)\,,
\eeqa
for $\Omega_B \gg \omega$.
In the opposite regime $z \ll 1$,
the singularity is within the effective region
of integration and requires some care for finite $z$.
In the extreme limit,
\beq
F_n(0) = (n - 2)!
\eeq
and
\beqa
\label{piTPerclassweak} \pi_T & = & \frac{e^2\beta}{3}(n_e +
n_{\bar e})\,,\nonumber\\ \pi_P & = & \frac{e^2\beta}{6}(n_e -
n_{\bar e})\frac{\Omega_B}{\omega} \,,
\eeqa
for $\Omega_B \ll \omega$. These expressions correspond to
the weak-field limit used in Ref.\ \cite{paletal}.

For arbitrary values of $z$,
the functions can be calculated by substituting
the equalities
\beqa
\frac{x^2}{x^2-z^2} & = & 1 + \frac{z}{2}\left(\frac{1}{x-z}
- \frac{1}{x+z}\right)\,,\nonumber\\
\frac{x^3}{x^2-z^2} & = & x +
\frac{z^2}{2}\left(\frac{1}{x-z} + \frac{1}{x+z}\right)\,,
\eeqa
in Eq.\ (\ref{Fn}), followed by the
change of variable $t = x - z$. In this way, $F_2$ and $F_3$
are found to be given by
\beqa
F_2(z) & = & 1 - \frac{z}{2}\left[e^{-z} Ei(z) - e^z Ei(-z)\right]\,,
\nonumber\\
F_3(z) & = & 1 - \frac{z^2}{2}
\left[e^{-z} Ei(z) + e^z Ei(-z)\right]\,,
\eeqa
in terms of the Exponential Integral function
\beq
Ei(z)= {\cal P}\int_{-\infty}^z \frac{e^{t}}{t} dt \,.
\eeq
By means of Eq.\ (\ref{piTPerclass1}),
$\pi_{T,P}$ can then be computed for any
value of $z$. In particular, using the expansion
\beq
Ei(z) = \ln(|z|) + \gamma + z + \frac{z^2}{4} + \frac{z^3}{18} + O(z^4) \,,
\eeq
where $\gamma=0.577215$ is the Euler constant, for $z$ small we
obtain
\beqa F_2(z) & = & 1 - z^2 + z^2(\gamma + \ln|z|) + O(z^4)\,,
\nonumber\\ F_3(z) & = & 1 - z^2 (\ln(|z|) + \gamma) + O(z^4) \,.
\eeqa
When these expressions are substituted in Eq.\ (\ref{piTPerclass1})
they of course reproduce the lowest order
terms given in Eq.\ (\ref{piTPerclassweak}) plus the
$O(\Omega^2_B)$ corrections to them.

\subsubsection{Case II - Degenerate gas}
This case corresponds to the limit
\beq
\mu \gg \frac{1}{\beta} \,.
\eeq
The distribution functions are
\beqa
f_e & = & \Theta(p_F - p) \,,\nonumber\\
f_{\bar e} & = & 0 \,,
\eeqa
where the Fermi momentum is the same as the chemical potential,
and its relationship with the electron number density is
\beq
n_e = \frac{p_F^3}{3\pi^2} \,.
\eeq

The formula for $\pi_L$ in Eq.\ (\ref{piTLPnsmallwrel}) yields
\beq
\pi_L = \frac{e^2 n_e}{p_F} \,,
\eeq
while the formulas for $\pi_{T,P}$ become
\beqa
\label {piTPrdeg}
\pi_T & = &
\left(\frac{e^2}{\pi^2}\right)\frac{2}{3}\int_{0}^{p_F} dp \frac{p^3}
{p^2 - \frac{e^2 B^2}{\omega^2}}\,, \nonumber\\[12pt]
\pi_P & = &
\left(\frac{e^2}{\pi^2}\right)\left(\frac{|e|B}{\omega}\right)\frac{1}{3}
\int_{0}^{p_F} dp \frac{p^2}{p^2 - \frac{e^2 B^2}{\omega^2}} \,.
\eeqa
With the instruction that the integrals in Eq.\ (\ref {piTPrdeg})
are to be evaluated by taking the principal value part, they yield
\beqa
\label{piTPreldeg}
\pi_T & = & \left(\frac{e^2 n_e}{p_F}\right)
\left(1 + \frac{\Omega_F^2}{\omega^2}
\log\left|1 - \frac{\omega^2}{\Omega_F^2}\right|\right)\,,\nonumber\\[12pt]
\pi_P & = & \left(\frac{e^2 n_e}{p_F}\right)
\frac{\Omega_F}{\omega} \left( 1 +
\frac{\Omega_F}{2\omega}\log\left|\frac{\omega - \Omega_F}
{\omega + \Omega_F}\right|\right)\,,
\eeqa
where
\beq
\Omega_F \equiv \frac{|e|B}{p_F} \,.
\eeq

\subsection{Weak-field limit}
\label{subsec:weakfieldlimit}
By this we mean that $\sqrt{|e|B}$ is sufficiently small, compared
to an average electron energy $\langle E_e\rangle$ and $\omega$,
so that we can literally take the $B \rightarrow 0$ limit
in Eqs.\ (\ref{piTLPgeneral}) and (\ref{piTLPn}), and keep
only the term that is linear in $B$. In this case,
Eqs.\ (\ref{relcasesmallB}), (\ref{smallB1}) and (\ref{smallB2})
hold. As shown in Appendix\ \ref{app:weakfieldlimit},
using them in Eq.\ (\ref{piTLPgeneral}) and taking
the $B\rightarrow 0$ limit results in
\beqa
\label{piTLPweakfield}
\pi_T{(\omega)}
& = & 4|e|^2\int\frac{d^3p}{(2\pi)^2 2E}
(f_e(E) + f_{\bar e}(E)) \frac{E^2 - \frac{1}{3}\vec p^2}
{E^2 - \omega^2/4}\,,\nonumber\\
\pi_P{(\omega)}
& = & 2|e|^2 \left(\frac{2|e|B}{\omega}\right)\int\frac{d^3p}{(2\pi)^2 2E}
(f_e(E) - f_{\bar e}(E))
\frac{E - \frac{2\vec p^{\;2}}{3E}}{E^2 - \omega^2/4}\,,\nonumber\\
\pi_L{(\omega)}
& = & 4|e|^2\int\frac{d^3p}{(2\pi)^2 2E}
(f_e(E) + f_{\bar e}(E))\frac{E^2 - \frac{1}{3}\vec p^2}{E^2 - \omega^2/4}\,,
\eeqa
where we have put
$\vec p_\perp^{\;2}\rightarrow \frac{2}{3}\vec p^{\;2}$ in the integrand.
For $\omega \ll 2\langle E_e\rangle$, these reduce further to
\beqa
\label{piTLweakfield}
\pi_T(\omega) = \pi_L(\omega) =
& = & 4|e|^2\int\frac{d^3p}{(2\pi)^2 2E}
(f_e(E) + f_{\bar e}(E))\left(1 - \frac{\vec p^2}{3E^2}\right) \,,
\eeqa
which is the standard expression for the plasma frequency squared,
and
\beq
\label{piPweakfield}
\pi_P(\omega) =  \left(\frac{4|e|^3 B}{\omega}\right)
\int\frac{d^3p}{(2\pi)^2 2E} \frac{(f_e(E) - f_{\bar e}(E))}{E}
\left(1 - \frac{2\vec p^{\;2}}{3E^2}\right) \,,
\eeq
both of which can be evaluated explicitly for the various limiting forms
of the electron distribution functions that we have considered.
We do not proceed any further in this direction, but we
mention that the results thus obtained agree with the corresponding ones
obtained in the previous sections, in their common range of validity.
That is, for example,
for a degenerate distribution in the relativistic regime,
from Eq.\ (\ref{piPweakfield}) we obtain
\beq
\pi_P = \frac{|e|^3B p_F}{3\pi^2\omega}\,,
\eeq
which agrees with the formula that is obtained from Eq.\ (\ref{piTPreldeg})
in the limit $\Omega_F \ll \omega$ (which corresponds to the weak-field
limit there). For a non-relativistic gas, putting $f_{\bar e} \approx 0$
and neglecting the term $\vec p^{\;2}/E^2$ in the integrand
of Eqs.\ (\ref{piTLweakfield}) and (\ref{piPweakfield}),
\beq
\pi_P(\omega) = \frac{\omega_B}{\omega}\pi_T(\omega)\,,
\eeq
with $\pi_T = \Omega^2_0$,
which coincides with the result obtained by taking the
weak-field limit ($\omega_B \ll \omega$) in Eq.\ (\ref{piTLPsmallwnr}).
Analogous relations can be verified for other cases as well.
However, we emphasize that the results of the previous sections
hold for a wider range of conditions than those for which
Eq.\ (\ref{piTLPweakfield}), and the relations based on it, hold.
The latter are limited
to situations in which retaining only the linear terms in $B$
is a valid approximation.

\subsection{Faraday Rotation}

The fact that the two transverse modes have different dispersion
relations leads to the Faraday rotation effect, as is well known.
After traveling a distance $L$, the direction of polarization
of the wave has rotated by an angle
\beq
\label{faradaytheta}
\theta = \frac{1}{2}\omega \Delta n L \,,
\eeq
where
\beq
\Delta n = (n_{-}(\omega) - n_{+}(\omega)) \,,
\eeq
with $n_{\mp}(\omega)$ being the refractive indices of the left
and right polarized modes, respectively. These can be computed
from the dispersion relations given above by using $n(\omega)=
Q/\omega$. For illustrative purposes, taking as an example the
case of a non-relativistic gas, for propagation parallel to the
magnetic field  from Eq.\ (\ref{disprelexample}) we find
\beqa \label{refractiveindex}
n^2_+(\omega)&=&1-\frac{\Omega^2_0}{\omega
(\omega-\omega_B)},\nonumber\\
n^2_-(\omega)&=&1-\frac{\Omega^2_0}{\omega (\omega+\omega_B)},
\eeqa
and therefore
\beq n^2_-(\omega)-n^2_+(\omega) =
\frac{2\Omega^2_0\omega_B}{\omega (\omega^2-\omega^2_B)}. \eeq
Writing
\beq
\Delta n =  \frac{n^2_-(\omega)-n^2_+(\omega)}{n_-(\omega)+n_+(\omega)} \,,
\eeq
we then obtain
\beq
\label{faraday}
\frac{\theta}{L} =\frac{\Omega^2_0\omega_B}{2n(\omega)(\omega^2-\omega^2_B)}\,,
\eeq
where we have defined $n(\omega) = (n_{-}(\omega) + n_{+}(\omega))/2$.

For values of $\omega$ such that $\omega \gg \Omega_0$,
Eq.\ (\ref{refractiveindex}) tells us that
$n_- \simeq n_+ \simeq 1$ and therefore
\beq
\label{faradayvacuum}
\frac{\theta}{L} \simeq \frac{\Omega^2_0\omega_B}{2(\omega^2-\omega^2_B)} \,.
\eeq
This result coincides with the one given in Ref.\cite{martin}. On the
other hand, for $\omega \gg \omega_B$, which corresponds to the 
weak-field approximation\footnote{This result agrees with what can be deduced
from the results given in Ref.\ \cite{paletal} for the 
non-relativistic gas. Notice that the quantity 
denoted by $d\Phi/d\ell$ in that reference
corresponds to $\omega\Delta n$ in our notation, and is twice the
rotation angle.},
\beq
\label{faradaylinear}
\frac{\theta}{L} \simeq \frac{\Omega^2_0\omega_B}{2n\omega^2} \,,
\eeq
where, from Eq.\ (\ref{refractiveindex}),
\beq
n \simeq \sqrt{1 -\frac{\Omega^2_0}{\omega^2}} \,.
\eeq
For other values of $\omega$, the formula in Eq.\ (\ref{faraday})
interpolates nicely between the two limiting cases we have
mentioned. 

More generally, for propagation parallel to the
magnetic field as we have considered above, the rotation
angle is given by Eq.\ (\ref{faradaytheta}) with
\beq
n^2_{\pm} = 1 - \frac{\pi_T \pm \pi_P}{\omega^2} \,.
\eeq
Thus, explicit formulas for the angle of rotation
can be deduced simply for the other situations of interest
using the results for $\pi_{T,P}$ that we have already obtained
for the various cases.
%
% sec 6
%
\section{Discussion and Conclusions}
\label{sec:conclusions}

The subject of the propagation of a photon in a background of
particles embedded in an external magnetic field appears in many
physical contexts. As we have mentioned in the Introduction, there
exist situations of interest for which neither the semiclassical
methods nor the linear field-theoretic approach are directly
applicable. In these cases, a more general field-theoretical
treatment that does not involve the weak field assumption is
required.

With this motivation in mind, in this work we have reexamined the
subject. We have given a general decomposition of the photon
self-energy in a matter background that contains a magnetic field,
in terms of the minimal set of tensors consistent with isotropy
and the transversality condition. From this result, we have shown
that the self-energy can be expressed in terms of nine independent
form factors, that in the long wavelength limit reduce to three.
In this limit, by applying the (real-time) finite temperature
field theory method, we have calculated the one-loop formulas
for the form factors. The formulas obtained in this way are valid
for arbitrary distributions of the electron gas and for
strong magnetic fields. They were explicitly evaluated for a
variety of conditions of interest in physical applications,
including the weak-field limit as a special case. 
From them we determined the photon dispersion relations and computed 
the Faraday rotation for various cases. 
They reproduce the well-known
semi-classical results when the appropriate limits are taken, but they
remain valid for more general situations, including those in which
the magnetic field is not weak.

\appendix
\section{Thermal electron propagator}
\label{app:propagator}

In this section we explain briefly the formula for the electron
propagator that will be used in the subsequent calculations.
Our staring point is the expression given in the book
by Itzykson and Zuber\cite{iz}. In vacuum, but in the presence of a constant
magnetic field, the electron propagator in coordinate space is
given by
\beq
\label{defprop}
S_A(x,x^\prime) = [i\lslash{\partial}_x - e \lslash{A}(x) + m](-i)
\int_{-\infty}^0 d\tau U(x,x^\prime;\tau) \,,
\eeq
where %
\footnote{
Our formula for $U$ differs from the one quoted in the book cited, 
by a factor of $e$ in the term $\frac{i}{2}\sigma\cdot F\tau$, which
we believe that it is missing there, and in the overall sign,
which believe is an error in the determination of the
normalization constant $C$ in the book.}  
\beqa
\label{Utau}
U(x,x^\prime;\tau) & = & \frac{i}{16\pi^2\tau^2}\phi(x,x^\prime)
\exp\left\{\frac{i}{4}(x - x^\prime)eF\coth(eF\tau)(x - x^\prime) +
i\left(\frac{e}{2}\sigma\cdot F + m^2 - i\epsilon\right)\tau\right.\nonumber\\
& & \mbox{} 
- \left.\frac{1}{2}\mbox{Tr}\,\ln\left[(eF\tau)^{-1}\sinh(eF\tau)\right]
\right\}\,,
\eeqa
and with the convention that $e$ is the charge of the electron.
The factor $\phi(x,x^\prime)$ is given in general by
\beq
\phi(x,x^\prime) = \exp\left\{-ie\int_{x^\prime}^{x}
d\xi^\mu \left[A_\mu(\xi) + 
\frac{1}{2}F_{\mu\nu}(\xi - x^\prime)^\nu
\right]\right\} \,,
\eeq
and it depends on the gauge but is independent of the path of 
integration,
%\footnote{The statement made in the Ref. \cite{kimetal}
%in this regard is incorrect.}, 
and it has the form
\beq
\label{phase}
\phi(x,x^\prime) = \exp\left\{\frac{ie}{2}F_{\mu\nu}x^\mu x^{\prime\,\nu}
\right\}\,,
\eeq
in the gauge in which
\beq
\label{gauge}
A_\mu = -\frac{1}{2}F_{\mu\nu}x^\nu \,.
\eeq
The expression for the propagator in momentum space is obtained 
by using the formula
\beq
\label{fourier}
e^{\frac{i}{4}(x - x^\prime)_\mu A^\mu{}_\nu (x - x^\prime)^\nu} = 
\left(\frac{-i}{\pi^2\sqrt{\mbox{Det}(A)}}\right)
\int d^4p e^{-i\left[
p_\mu(A^{-1})^\mu{}_\nu p^\nu + (x - x^\prime)\cdot p
\right]} \,,
\eeq
where $\mbox{Det}(A)$ stands for the determinant of the matrix
formed by the coefficients $A^\mu{}_\nu$.
For the ensuing manipulations it is useful to observe that
\beq
F_{\mu\nu} = iB P_{\mu\nu}\,,
\eeq
where $P_{\mu\nu}$ is defined in Eq.\ (\ref{deftensors}).  Using the
multiplication rules $P^2 = R$ and $R^2 = R$, the following 
relations are readily derived,
\beqa
\cosh(eF\tau) & = & 1 + R\left[\cos(s) - 1\right]\,,\nonumber\\
\left[\cosh(eF\tau)\right]^{-1} & = & 
1 + R\left[\frac{1}{cos(s)} - 1\right] \,,\nonumber\\ 
(eF\tau)^{-1}\sinh(eF\tau) & = & 1 + R\left[\frac{\sin(s)}{s} - 1\right] \,,
\eeqa
from which the rest of the formulas that we need can be obtained. 
In particular,
\beqa
\label{matrixidentities}
\left[(eF\tau)\coth(eF\tau)\right]^{-1} & = & 1 + 
R\left[\frac{\tan(s)}{s} - 1\right]\,,\nonumber\\
\mbox{Det}\left(\left[(eF\tau)\coth(eF\tau)\right]^{-1}\right) & = &
\left(\frac{\tan(s)}{s}\right)^2 \,,\nonumber\\
\mbox{Tr}\ln\left[(eF\tau)^{-1}\sinh(eF\tau)\right] & = & 
\ln\left(\frac{\sin(s)}{s}\right)^2 \,,
\eeqa
where $s$ is the variable defined in Eq.\ (\ref{vars}).

Thus, using Eqs.\ (\ref{fourier}) in Eq. (\ref{Utau}), 
with the help of Eq.\  (\ref{matrixidentities}) we obtain
\beq
\label{Utrans}
U(x,x^\prime;\tau) =
\phi\frac{1}{\cos(s)}\int\frac{d^4p}{(2\pi)^4}e^{-ip\cdot(x - x^\prime)}
e^{-i\tau\left[\Phi(p,s) - \frac{e}{2}\sigma\cdot F + i\epsilon\right]} \,,
\eeq
with $\Phi$ as defined in Eq.\ (\ref{GPhi}) in the text. 
For the gauge choice given in Eq.\ (\ref{gauge}), which implies
Eq.\ (\ref{phase}), it follows that
\beq
(i\lslash{\partial}_x - e\lslash{A} + m)U(x,x^\prime;\tau) = 
\phi\frac{1}{\cos(s)}\int\frac{d^4p}{(2\pi)^4}e^{-ip\cdot(x - x^\prime)}
\left[\lslash{p} - \tan(s)iP_{\mu\nu}\gamma^\mu p^\nu + m\right]
e^{-i\tau\left[\Phi(p,s) - \frac{e}{2}\sigma\cdot F + i\epsilon\right]}\,.
\eeq
Using the relation
\beq
e^{\frac{ie\tau}{2}\sigma\cdot F} =  
\cos(eB\tau) + i\gamma_5\lslash{u}\lslash{b}\sin(eB\tau)\,,
\eeq
which follows by noticing that
\beq
\frac{1}{2}\sigma\cdot F = B\gamma_5\lslash{u}\lslash{b}\,,
\eeq
together with the fact that $(\gamma_5\lslash{u}\lslash{b})^2$ = 1,
Eq.\ (\ref{defprop}) then yields
\beq
\label{propx}
S_A(x,x^\prime) = 
\phi\int\frac{d^4p}{(2\pi)^4}e^{-ip\cdot(x - x^\prime)}S_F(k) \,,
\eeq
where
\beq
\label{SFaux}
iS_F(p) = \int_0^{\infty} d\tau G(p,s)e^{i\tau\Phi(p,s) - \epsilon\tau} \,,
\eeq
with
\beq
\label{G1}
G(p,s) = \left[\lslash{p} - \tan(s)iP_{\mu\nu}\gamma^\mu p^\nu + m\right]
\left[1 +i\gamma_5\lslash{u}\lslash{b}\tan(s)\right]\,.
\eeq
In writing Eq.\ (\ref{SFaux}), we have made the change of variable 
$\tau \rightarrow -\tau$. Using the identity
\beq
P_{\mu\nu}\gamma^\mu k^\nu = -\gamma_5\lslash{k}\lslash{u}\lslash{b} +
(k\cdot u)\gamma_5\lslash{b} - (k\cdot b)\gamma_5\lslash{u} \,,
\eeq
$G$ can be expressed in the final form given in Eq.\ (\ref{GPhi}) in the text.
Finally, the 11 component of the 
thermal propagator is obtained from the formula
\beq
S_e(p) = S_F(p) - \left(S_F(p) - \bar S_F(p)\right)\eta_e \,,
\eeq
where $\eta_e$ is the background-dependent factor defined in
Eq.\ (\ref{etae}) and, as usual, $\bar S_F = \gamma^0 S_F^\dagger\gamma^0$.
This leads to the decomposition of $S_e$ given in Eq.\ (\ref{Se}),
with the thermal part $S_{Te}$ as defined in Eq.\ (\ref{STe}).

We would like to mention the following point.  Strictly,
the Fourier expansion given in Eq.\ (\ref{Utrans}) holds
only for those values of $\tau$ for which $\cos(s) \not = 0$.
At the points where $\cos(s) = 0$, the expansion given
in Eq.\ (\ref{Utrans}) is not valid, and the integration over $\tau$ in
Eq.\ (\ref{SFaux}) is not defined at those points as it stands.
Therefore, we \emph{define} the $\tau$ integration in Eq.\ (\ref{SFaux}),
by deforming the line of integration to lie just below the real axis,
so that the points in question are avoided.  It is not difficult to see that
this choice, as well any other that avoids
those points (such as, for example, choosing
the integration line to lie just above the real axis), 
amounts to a redefinition of the function $\phi(x,x^\prime)$.

\section{Evaluation of $J_0$}
\label{app:J0}

In order to evaluate the integral $J_0$ defined in
Eq.\ (\ref{Ji}), we first observe that, 
as a function of $\psi$, $J_0$ is analytic in the upper-half plane. 
Therefore, if we can evaluate it for a particular subregion of that
plane, the result is valid for the entire region by analytic
continuation.  For the subregion $\mbox{Re}\,\psi < 0$ 
(or, equivalently, Euclidean momenta $p_{||}$) we proceed
as folows.

The integrand is an analytic function of $s$ in the lower-half $s$-plane. 
Therefore we can deform the integration path and integrate over the
negative imaginary axis. Then setting
\begin{equation}
s = -it \,,
\end{equation}
we have
\begin{equation}
\label{J1aux}
J_0 = \frac{-i}{|e|B}
\int_0^\infty dt e^{t\psi} e^{-a} e^{\frac{2au}{1 + u}} \,,
\end{equation}
where
\begin{equation}
u \equiv e^{-2t} \,,
\end{equation}
and we have used
\begin{equation}
-i\tan(-it) = \frac{2u}{1 + u} - 1 \,.
\end{equation}
Using the generating formula of the Legerre polynomials
\begin{equation}
\frac{e^{-\frac{xz}{1 - z}}}{1 - z} = \sum_{n = 0}^\infty L_n(x)z^n\,,
\end{equation}
(valid for $0 \le x < \infty$ and $|z| < 1$),
equation\ (\ref{J1aux}) becomes
\begin{equation}
\label{J1aux2}
J_0 = \frac{-i}{|e|B}
\int_0^\infty dt e^{t\psi}\sum_{n = 0}^\infty D_n(a)e^{-2tn} \,,
\end{equation}
where $D_n(a)$ is the function defined in Eq.\ (\ref{Dn}) in the text.
Equation\ (\ref{J1aux2}) is easily integrated term by term to yield
the final result  quoted in Eq.\ (\ref{J1final}),
which also holds for $\mbox{Re\,}\psi > 0$ (or equivalently, Minkowskian
values of $p_{||}$) by analytic continuation.

%
% appendix 3
%
\section{Low Frequency limit}.
\label{app:lowfrequencyregime}

To arrive at Eqs.\ (\ref{piTPnsmallw}) and (\ref{piLnsmallw})
we apply the low frequency conditions to Eq.\ (\ref{piTLPn}) as follows.
Let us define
\beq
\label{D}
D_{\pm} \equiv \frac{1}{\omega^2 + 2\omega E_n - 2|e|B} \pm
\frac{1}{\omega^2 - 2\omega E_n - 2|e|B} \,.
\eeq
By combining the denominators,
\beqa
\label{Dexact}
D_{+} & = &
\frac{2(\omega^2 - 2|e|B)}{(\omega^2 - 2|e|B)^2 - 4\omega^2 E^2_n}\,,
\nonumber\\
D_{-} & = &
\frac{-4\omega E_n}{(\omega^2 - 2|e|B)^2 - 4\omega^2 E^2_n} \,.
\eeqa
Now define the combinations
\beqa \label{Delta}
\Delta^{(\pm)}_{+} & = & D_{+} \pm (B
\rightarrow -B)\,,\nonumber\\
\Delta^{(\pm)}_{-} & = & D_{-} \pm
(B \rightarrow -B)\,, \eeqa
in terms of which the formulas for $\pi^{(n)}_{T,P}$ given
in Eq.\ (\ref{piTLPn}) are expressed
as
\beqa
\label{piTLPintermsofdelta}
\pi^{(n\not= 0)}_T{(\omega)} & = & \int_{-\infty}^\infty
\frac{dp_{\parallel}}{E_n} (f_e(E_n) + f_{\bar e}(E_n))
\left\{
\omega E_n\Delta^{(+)}_{-} + 2n|e|B\Delta^{(+)}_{+}\right\}\,,\nonumber\\
\pi^{(0)}_T{(\omega)} & = & \int_{-\infty}^\infty
dp_{\parallel} (f_e(E_n) + f_{\bar e}(E_n))
\left\{\omega D_{-}\right\}\,,\nonumber\\
\pi^{(n\not= 0)}_P{(\omega)} & = & \int_{-\infty}^\infty
\frac{dp_{\parallel}}{E_n} (f_e(E_n) - f_{\bar e}(E_n))
\left\{
\omega E_n\Delta^{(-)}_{+} + 2n|e|B\Delta^{(-)}_{-}\right\}\,,\nonumber\\
\pi^{(0)}_P{(\omega)} & = & \int_{-\infty}^\infty
dp_{\parallel} (f_e(E_n) - f_{\bar e}(E_n))
\left\{\omega D_{+}\right\} \,.
\eeqa
{From} their definitions, and using Eq.\ (\ref{D}),
\beqa \label{Deltaexact} \Delta^{(\pm)}_{+} & = & \frac{2(\omega^2
- 2|e|B)}{(\omega^2 - 2|e|B)^2 - 4\omega^2 E^2_n} \pm
\frac{2(\omega^2 + 2|e|B)}{(\omega^2 + 2|e|B)^2 - 4\omega^2
E^2_n}\,, \nonumber\\
\Delta^{(\pm)}_{-} & = & \frac{-4\omega E_n}{(\omega^2 - 2|e|B)^2
- 4\omega^2 E^2_n} \pm \frac{-4\omega E_n}{(\omega^2 + 2|e|B)^2 -
4\omega^2 E^2_n} \,.
\eeqa
It is seen that all the $\Delta$'s can be
approximated by letting $(\omega^2 - 2|e|B)^2 \rightarrow 4|e|^2 B^2$
in the denominators, except $\Delta^{(-)}_{-}$, which would give zero.
This in particular means that we cannot take the low frequency limit
in Eq.\ (\ref{piTLPn}) by just setting $\omega^2 \rightarrow 0$
in the denominators, or otherwise we would miss some important terms.
In order to make a systematic expansion of $\Delta^{(-)}_{-}$
for $\omega \ll \sqrt{|e|B}$, we use the relation
\beq
D_{-} = \frac{1}{\omega E_n} - \frac{\omega^2 - 2|e|B}{2\omega E_n}D_{+} \,,
\eeq
which follows from using the trivial identity
\beq \frac{1}{x + y} - \frac{1}{x - y} = \frac{1}{y} \left[2 -
x\left(\frac{1}{x + y} + \frac{1}{x - y}\right)\right]\,. \eeq
Then going back to the definition in Eq.\ (\ref{Delta}),
we can re-express $\Delta^{(-)}_{-}$ as
\beq
\label{Delta--}
\Delta^{(-)}_{-} = -\left(\frac{\omega}{2E_n}\right)\Delta^{(-)}_{+}
+ \left(\frac{|e|B}{\omega E_n}\right)\Delta^{(+)}_{+} \,.
\eeq
Up to this point the relations are exact. Now we can proceed to
make the approximation $\omega \ll \sqrt{|e|B}$ systematically,
by letting $(\omega^2 \pm 2|e|B) \rightarrow \pm 2|e|B$
in Eq.\ (\ref{Deltaexact}), except for
$\Delta^{(-)}_{-}$, for which we use Eq.\ (\ref{Delta--}).
Thus,
\beqa \Delta^{(+)}_{+} & = & \frac{-\omega^2}{\omega^2 E^2_n -
|e|^2 B^2}\,, \nonumber\\[12pt] \Delta^{(-)}_{+} & =
&\frac{2|e|B}{\omega^2 E^2_n - |e|^2 B^2}\,, \nonumber\\[12pt]
\Delta^{(+)}_{-} & = & \frac{2\omega E_n}{\omega^2 E^2_n - |e|^2
B^2}\,, \nonumber\\[12pt] \Delta^{(-)}_{-} & = &
\left(\frac{-2\omega |e|B}{E_n}\right) \frac{1}{\omega^2 E^2_n -
|e|^2 B^2} \,. \eeqa
Similarly, from Eq.\ (\ref{Dexact})
\beqa
D_{+} & = & \frac{|e|B)}{\omega^2 E^2_n - |e|^2 B^2}
\nonumber\\
D_{-} & = & \frac{\omega E_n}{\omega^2 E^2_n - |e|^2 B^2}\,.
\eeqa
The formulas quoted in Eq.\ (\ref{piTPnsmallw})
follow by substituting these approximate expressions for the $\Delta$'s
and the $D$'s into Eq.\ (\ref{piTLPintermsofdelta}).

For $\pi^{(n)}_L{(\omega)}$ we assume that $\omega \ll 2\langle E_e\rangle$
to approximate
\beq
\frac{1}{\omega^2 + 2\omega E_n} + \frac{1}{\omega^2 - 2\omega E_n}
\rightarrow \frac{-1}{2E^2_n} \,,
\eeq
from which the formula in Eq.\ (\ref{piLnsmallw}) follows.
%
% appendix 4
%
\section{The weak-field (linear) limit}
\label{app:weakfieldlimit}

In order to obtain the weak-field limit formulas
quoted in Eq.\ (\ref{piTLPweakfield}),
we use the expressions given in Eq.\ (\ref{piTLPintermsofdelta})
for $\pi_{T,P}(\omega)$. From Eq.\ (\ref{Deltaexact}),
as $B \rightarrow 0$
\beqa
\Delta^{(+)}_{-} & \rightarrow &
\frac{2\omega E}{\omega^2(E^2 - \omega^2/4)}\,,
\nonumber\\
\Delta^{(+)}_{+} & \rightarrow & \frac{-1}{E^2 - \omega^2/4} \,.
\eeqa
Then, remembering that Eq.\ (\ref{smallB2}) holds,
\beqa
\pi_T{(\omega)} & = &
2|e|^2\int\frac{d^3p}{(2\pi)^2 2E}
(f_e(E) + f_{\bar e}(E))\left(\frac{1}{E^2 - \omega^2/4}\right)
\left[2E^2 - \vec p_\perp^{\;2}\right] \,,
%\nonumber\\
%& = & 4|e|^2\int\frac{d^3p}{(2\pi)^2 2E}
%(f_e(E) + f_{\bar e}(E))\left(\frac{E^2}{E^2 - \omega^2/4}\right)
%\left[1 - \frac{\vec p^{\;2}}{3E^2}\right] \,,
\eeqa
which yields the formula given in Eq.\ (\ref{piTLPweakfield}),
after replacing $\vec p_\perp^{\;2}\rightarrow \frac{2}{3}\vec p^{\;2}$
in the integrand. We proceed similarly for $\pi_P$.
{From} Eq.\ (\ref{Deltaexact}),
\beqa
\Delta^{(-)}_{+} & \rightarrow & \frac{-8|e|B}{\omega^4 - 4\omega^2 E^2}\,,
\nonumber\\
\Delta^{(-)}_{-} & \rightarrow &
-\left(\frac{\omega^2}{2\omega E}\right) \Delta^{(-)}_{+} +
\frac{|e|B}{\omega E}\Delta^{(+)}_{+} \nonumber\\
& = & \frac{|e|B}{\omega E}\frac{8\omega^2}{\omega^4 - 4E^2\omega^2} \,,
\eeqa
where we have used Eq.\ (\ref{Delta--}) for
$\Delta^{(-)}_{-}$.
Then substituting in Eq.\ (\ref{piTLPintermsofdelta}),
\beqa \pi_P(\omega) & = & 2|e|^2 \left(\frac{2|e|B}{\omega}\right)
\int\frac{d^3p}{(2\pi)^2 2E} (f_e(E) - f_{\bar e}(E)) \frac{E -
\frac{\vec p_\perp^{\;2}}{E}}{E^2 - \omega^2/4} \,.
%& = & \left(\frac{4|e|^3B}{\omega}\right)
%\int\frac{d^3p}{(2\pi)^2 2E} (f_e(E) - f_{\bar e}(E))
%\frac{E - \frac{2\vec p^{\;2}}{3E}}{E^2 - \omega^2/4} \,.
\eeqa
Finally, for $\pi_L(\omega)$ we apply Eq.\ (\ref{smallB2})
directly in Eqs.\ (\ref{piTLPgeneral}) and (\ref{piTLPn}) and obtain
\beqa
\pi_L{(\omega)} & = & 4|e|^2\int\frac{d^3p}{(2\pi)^2 2E}
(f_e(E) + f_{\bar e}(E))[4m_e^2 + 4\vec p_\perp^{\,2}]
\frac{1}{4E^2 - \omega^2} \,,
%\nonumber\\
%& = & 4|e|^2\int\frac{d^3p}{(2\pi)^2 2E}
%(f_e(E) + f_{\bar e}(E))\frac{E^2 - \frac{1}{3}\vec p^2}{E^2 - \omega^2/4} \,,
\eeqa
which can be expressed as quoted in Eq.\ (\ref{piTLPweakfield}).

\end{document}